\documentclass[manuscript]{aastex61}
\pdfoutput=1 %for arXiv submission
\usepackage{amsmath,amstext}
\usepackage[T1]{fontenc}
\usepackage{apjfonts} 
\usepackage[figure,figure*]{hypcap}
\usepackage{url} 
\usepackage{amsmath}
\usepackage{booktabs}
\usepackage{hyperref}
\usepackage{subfigure}
\usepackage{enumerate}
\usepackage{comment}
\usepackage{xspace}
\usepackage{soul}
\usepackage{threeparttable}
\usepackage{graphicx}

\newcommand{\teff}{$T_\mathrm{eff}$ }

\newcommand{\Rearth}{R$_\oplus$\xspace}
\newcommand{\Mearth}{M$_\oplus$\xspace}

\shorttitle{PFS RV Masses of GJ 9827 Planets}
\shortauthors{Teske et al.}

\begin{document}

%\title{Magellan II/PFS Radial Velocity Mass Measurements of the Super-Earth Planets Transiting GJ 9827 at 30 Parsecs}
\title{Magellan/PFS Radial Velocities of GJ 9827, a late K dwarf at 30 pc with Three Transiting Super-Earths}

\author{Johanna K. Teske}
\altaffiliation{Carnegie Origins Fellow, jointly appointed by Carnegie DTM and Observatories} 
\affiliation{Observatories of the Carnegie Institution for Science, 813 Santa Barbara St., Pasadena, CA 91101}
\affiliation{Department of Terrestrial Magnetism, Carnegie Institution for Science, 5241 Broad Branch Road, NW, Washington, DC 20015}
\author{Sharon Wang}
\affiliation{Department of Terrestrial Magnetism, Carnegie Institution for Science, 5241 Broad Branch Road, NW, Washington, DC 20015}
\author{Angie Wolfgang}
\altaffiliation{NSF Astronomy \& Astrophysics Postdoctoral Fellow}
\affiliation{Center for Exoplanets and Habitable Worlds, 525 Davey Laboratory, The Pennsylvania State University, University Park, PA 16802, USA}
\affiliation{Department of Astronomy and Astrophysics, The Pennsylvania State University, 525 Davey Laboratory, University Park, PA 16802, USA}
\author{Fei Dai}
\affiliation{Department of Physics and Kavli Institute for Astrophysics and Space Research, Massachusetts Institute of Technology, Cambridge, MA, 02139, USA}
\affiliation{Department of Astrophysical Sciences, Princeton University, 4 Ivy Lane, Princeton, NJ, 08544, USA}
\author{Stephen A. Shectman}
\affiliation{Observatories of the Carnegie Institution for Science, 813 Santa Barbara St., Pasadena, CA 91101}
\author{R. Paul Butler}
\affiliation{Department of Terrestrial Magnetism, Carnegie Institution for Science, 5241 Broad Branch Road, NW, Washington, DC 20015}
\author{Jeffrey D. Crane}
\affiliation{Observatories of the Carnegie Institution for Science, 813 Santa Barbara St., Pasadena, CA 91101}
\author{Ian B. Thompson}
\affiliation{Observatories of the Carnegie Institution for Science, 813 Santa Barbara St., Pasadena, CA 91101}

\accepted{January 25, 2018}

%%%%ABSTRACT%%%%
\begin{abstract}
The \textit{Kepler} mission showed us that planets with sizes between that of Earth and Neptune appear to be the most common type in our Galaxy. These ``super-Earths'' continue to be of great interest for exoplanet formation, evolution, and composition studies. However, the number of super-Earths with well-constrained mass and radius measurements remains small (40 planets with $\sigma_{\rm{mass}}<$ 25\%), due in part to the faintness of their host stars causing ground-based mass measurements to be challenging. Recently, three transiting super-Earth planets were detected by the \textit{K2} mission around the nearby star GJ 9827/HIP 115752, at only 30 pc away. The radii of the planets span the ``radius gap'' detected by Fulton et al. (2017), and all orbit within $\sim$6.5 days, easing follow-up observations. Here we report radial velocity (RV) observations of GJ 9827, taken between 2010 and 2016 with the Planet Finder Spectrograph on the Magellan II Telescope. We employ two different RV analysis packages, SYSTEMIC and \textsc{RadVel}, to derive masses and thus densities of the GJ 9827 planets. We also test a Gaussian Process regression analysis, but find the correlated stellar noise is not well constrained by the PFS data, and that the GP tends to over fit the RV semi-amplitudes resulting in a lower $K$ value. Our RV observations are not able to place strong mass constraints on the two outer planets (c \& d) but do indicate that planet b, at 1.64 R$_{\oplus}$ and $\sim8$ M$_{\oplus}$, is one of the most massive (and dense) super-Earth planets detected to date.  

\end{abstract}


%%%%SECTION I: Introduction%%%%

\section{Introduction}

One of the most profound results to come out of the \textit{Kepler} mission \citep{borucki2010,borucki2011,batalha2013,burke2014} is that the most common type of planet in the Galaxy, within orbital periods of $\sim$100 days, is between the size of Earth and Neptune ($\sim$1-4 R$_{\oplus}$), so called "super-Earth" exoplanets \citep[e.g.,][]{howard2012,batalha2013,petigura2013}. %This was unexpected based on previous models of planet formation, corroborated by the planets in our solar system (CITATION) and later the detected giant exoplanets (CITATION) that dominated the known exoplanet population for over a decade. 
Because super-Earths are so common and because we lack a (known) planet of this size in our solar system, understanding the formation mechanisms and compositions of super-Earth exoplanets is of great interest. In particular, the statistics of potentially ``Earth-like'' planets are affected by the nature of the smallest super-Earths (1-2 R$_{\oplus}$) -- are they more akin to the terrestrial planets in our solar system, dominated by rock, or do they instead have volatile envelopes representing a significant fraction of their radii?

This question was investigated as part of a large radial velocity (RV) follow-up program to observe the \textit{Kepler}-detected super-Earth planets. Through these efforts, a break in the mass-radius (M-R) distribution of planets was discovered around 1.6 R$_{\oplus}$, indicating that above this radius planets have a substantial gaseous atmosphere, and below this radius planets are mostly rocky \citep{rogers2015}. %Note that weiss&marcy2014 didn't discover this break, they assumed it based on Leslie's work.
Recently the California Kepler Survey (CKS) produced an updated catalog of 2025 \textit{Kepler} planet radii \citep{fulton2017}, using more precise stellar radii (from median uncertainties of 25\% in \citealt{huber2014} to 10\% in \citeauthor{fulton2017}) garnered from high resolution spectroscopy \citep{petigura2017} and isochrone fitting \citep{johnson2017}. After selecting a subset of these planet radii, focusing on the best characterized dwarf host stars with verified planets, \cite{fulton2017} found a gap in the radius distribution and thus planet occurrence rate distribution between $\sim$1.5-2 R$_{\oplus}$, coinciding with the M-R break previously observed. %While \cite{owen&wu2013} provided the first tentative evidence of this radius gap and were able to match the observations with hydrodynamical models of photoevaporation of planets' volatile envelopes, \cite{fulton2017}'s more precise planet radii measurements solidify the detection of the exoplanet radius gap.   AW: Owen & Wu's detection of a bimodality was due to a hidden selection effect in the stellar radius distribution that they didn't account for, not due to an intrinsic bimodality the planet radii (See Figure 1 of my 2012 paper for an illustration of this - the stars with Kp > 14 were systematically cooler and smaller than those with Kp<14, and so there were more small planets found around those stars).
Further modeling work by \cite{lopez2016}, \cite{owen&wu2017}, and \cite{jin&mordasini2017} has confirmed that the radius gap can be explained by evaporation, and suggest that super-Earth planets (at least within 100-d periods) probably form at or close to their current locations within water ice lines, and that they likely all start as $\sim$Earth-composition cores with a few percent weight of H/He envelopes (but which contribute to $\sim$half of the planet's radius). 

However, the question still remains: Can evaporation alone explain the astrophysical spread (e.g., \citealt{wolfgang2016}) in observed radii \textit{and} masses of super-Earth planets? \cite{owen&wu2017} specifically suggest that the radius gap can only be explained by $\lesssim$ a factor of two range in planet densities, corresponding to an iron fraction varying between 0-70\%. And yet we see a larger scatter than this in the known densities of super-Earths -- selecting the small planets with well-constrained radius and mass errors ($<1$ R$_{\oplus}$ and $<1.5$ M$_{\oplus}$, respectively) from the NASA Exoplanet Archive results in a range in density between 0.13$\times$ and 4.41$\times$ that of the Earth.\footnote{The mass range of this sample is 0.41-17.2 M$_{\oplus}$ with 0.85 M$_{\oplus}$ median errors, and the radius range of 0.77-2.99 R$_{\oplus}$ with 0.08 R$_{\oplus}$ median errors.} Even restricting this sample to only planets with 1.5 R$_{\oplus} < R_{p} < 2.0$  R$_{\oplus}$ still gives densities varying by a factor of 12. To help further test and refine the evaporation model for super-Earth planets, and thus better understand their formation and range (or lack thereof) in compositions, more measured super-Earth masses are required. 

Continuing the legacy of \textit{Kepler}, the \textit{K2} mission has provided new super-Earth planet detections, albeit often in shorter orbital periods due to the across-the-ecliptic pattern of \textit{K2}'s observations, versus the staring mode of its predecessor. \citet{niraula2017} and \citet{rodriguez2017} recently announced the \textit{K2} detection of a three-super-Earth system around a nearby (30 pc) K dwarf, GJ 9827, actually representing the closest exoplanet host discovered by the mission to date. Intriguingly, the radii of the three planets -- 1.64 (planet b), 1.29 (planet c), and 2.08 (planet d) R$_{\oplus}$ \citep{rodriguez2017} --  span the \cite{fulton2017} radius gap, and are all at relatively short orbital periods ($\lesssim 6.2$ days), facilitating both follow-up transit and RV observations. Coincidentally, GJ 9827 is a star that we have been monitoring with the Magellan II Planet Finder Spectrograph (detailed below) since its commissioning in 2010. Here we present our RV observations of GJ 9827 and the resulting mass constraints for each of the \textit{K2}-detected planets. In \S2 we outline the details of our observation. In \S3 we first present a derivation of stellar parameters from our high resolution, high signal-to-noise ``template'' (iodine-free) spectra, and then outline the derivation of the planet masses via two different RV fitting programs (SYSTEMIC and \textsc{RadVel}). We discuss the implications of our results for the nature of the GJ 9827 planets in \S4, and also list our conclusions there.

%%%%SECTION II: Observations%%%%
\section{Magellan/PFS Observations}
The radial velocity (RV) observations of GJ 9827 presented here were taken as part of the large Magellan Planet Search Program, a survey for planets orbiting close ($<$100 pc), quiet (little emission in the Ca II H \& K lines) FGKM dwarf stars that was started in 2002 using the MIKE echelle spectrograph \citep{bernstein2003}. In 2010 the survey switched to using a dedicated precision RV spectrograph, the Planet Finder Spectrograph (PFS, \cite{crane2006,crane2008,crane2010}), on the Magellan II (Clay) Telescope. Both MIKE and PFS contain an iodine absorption cell \citep{marcybutler1992} in the optical path that imprints on the incoming starlight a reference iodine spectrum, which is used to measure the instrument point spread function and for determining a precise wavelength solution. The stellar Doppler shifts are measured from the spectra using the technique described in \cite{butler1996}. In brief, the iodine region of the spectrum (between $\sim$5000-6200 {\AA}) is divided into 2 {\AA} chunks, on which a forward modeling procedure is performed, providing an independent
measurement of the instrument PSF, wavelength solution and Doppler shift. Part of this forward modeling also requires a high resolution, high-S/N ``template'', iodine-free spectrum of the star. In the case of GJ 9827, the template spectrum was observed through the 0.3x2.5$\arcsec$ slit on 21 August 2010 UT. The reported velocity for each observation (individual stellar spectrum) is then the weighted mean of the independent chunk velocities, and the internal uncertainty is the standard
deviation of all of the chunk velocities measured from that observation. The weighted mean Doppler shift and internal uncertainty for each observation is reported in \hyperref[tab:rvdata]{Table 1}. 

The thirty-six PFS observations of GJ 9827 span January 2010 to August 2016 and were taken through a 0.5x3.7$\arcsec$ or 0.5x2.5$\arcsec$ slit as noted in \hyperref[tab:rvdata]{Table 1}, providing a resolution of $\sim$80,000. The individual exposure times ranged from 457 s to 900 s; in the regular Magellan survey the data are usually binned on a nightly basis, but we left the data unbinned in this analysis due to the known short periods of the planets. PFS calibrations taken at the beginning of each night include 20-30 flat-field images, two iodine exposures, two rapidly rotating B star exposures, and one or two thorium argon exposures. Reduction of the raw CCD images and spectral extraction were carried out using a custom IDL-based pipeline that performs flat fielding, removes cosmic rays, and measures and subtracts scattered light. No sky subtraction is done, as our targets are all relatively bright.

We also include in \hyperref[tab:rvdata]{Table 1} a proxy for the $S$-index of each observation, which we measured as the ratio of the flux in the Ca H line at 3968.47\,{\AA} to the flux in an adjacent continuum region, centered at 3996.5\,{\AA} as in \citet{santos2001} (referred to as $S_{COR}$ in that work). The traditional $S$-index compares the flux in triangle-weighted bins with full width at half maximums of 1.09\,{\AA} centered on the Ca II H\&K lines (at 3968.47\,{\AA} and 3933.66\,{\AA}) to the flux in two rectangular 20\,{\AA}-wide continuum regions centered on 3901 and 4001\,{\AA} (Duncan et al. 1991). This index is known to be correlated with spot activity on the stellar surface, and serves as a proxy for chromospheric activity that could cause radial velocity shifts that mimic those induced by planets. 
 
\startlongtable
\begin{deluxetable}{cccc}
\tablecaption{PFS radial velocities and $S$-index values for GJ 9827 \label{tab:rvdata}}
\tablecolumns{4}
\tablewidth{0pc}
\tablehead{\colhead{BJD} & \colhead{RV [m\,s$^{-1}$]} & \colhead{Uncertainty [m\,s$^{-1}$]} & \colhead{$S$-index}}
\startdata
2455198.54192\tablenotemark{a}  &   12.27  &  1.59 &  0.5803 \\
2455198.55208\tablenotemark{a}  &    11.47 &   1.68 &  0.8042 \\
2455201.55456\tablenotemark{a}  &     6.75 &   3.16 &  0.7396 \\
2455201.56113\tablenotemark{a}  &    -8.28 &   3.78 &  0.5066 \\
2455428.80228  &    -6.69 &  1.49 &  0.5080 \\
2455429.85067  &    -7.03 &   1.58&  0.4946 \\
2455435.76931  &     0.59 &   1.46 &  0.5040 \\
2455437.78436  &     6.39 &   1.41 &  0.5132 \\
2455438.82524  &     7.51 &   1.50 &  0.5441 \\
2455439.82385  &     3.00 &   1.64 &  0.5476 \\
2455583.52551  &    -6.85 &   3.10 &  1.0763 \\
2455785.72196  &     2.42 &   1.56 &  0.7217 \\
2455786.69601  &    -7.28 &   1.52 &  0.0488 \\
2455787.64461  &     0.00 &   1.83 &  0.7111 \\
2455788.71617  &     0.04 &   1.86 &  0.0496 \\
2455790.72279  &     5.48 &   2.00 &  0.6268 \\
2455791.78779  &    -0.93 &   2.09 &  0.0498 \\
2455793.72693  &   -12.14 &   1.98 &  0.6180 \\
2455794.84266  &    -9.87 &   2.41 &  0.0494 \\
2455795.78402  &    -5.26 &   1.67&   0.5074 \\
2455843.77510  &    -5.96 &   2.05&   0.7916 \\
2455844.69159  &    -4.95 &   2.71 &  0.5659 \\
2455845.70111  &    -2.64 &   2.33 &  0.5983 \\
2456139.85610  &     5.05 &   1.66 &  0.6735 \\
2456147.80625  &    -7.97 &   3.02 &  0.7428 \\
2456150.82577  &    -0.96 &   1.81 &  0.6004 \\
2456509.86896  &    -4.33 &   1.60 &   0.6079 \\
2456603.67306  &    -1.40 &   1.90 &  0.6334 \\
2456610.60397  &     9.14 &   1.75 &  0.4721 \\
2456866.86057  &     8.22 &   1.91 &  0.7564 \\
2456877.85427  &     7.53 &   1.75 &  0.5226 \\
2457021.53642  &    10.17 &   1.52 &  0.7001 \\
2457203.88063  &    -2.43 &   2.02 &  0.6106 \\
2457261.84656  &     1.28 &   1.98&   0.6400 \\
2457319.65281  &     1.70 &   1.59 &  0.4304 \\
2457615.79012  &    -9.56 &   1.44 &  0.4988 \\
\enddata
\tablenotetext{a}{These observations were taken with a 0.5x3.7$\arcsec$ versus the normal 0.5x2.5$\arcsec$ slit.}
\end{deluxetable}

%%%%SECTION III: Results%%%%
\section{Analysis \& Results}
\subsection{Stellar Characterization }\label{sec:stellar}
\citet{niraula2017} reported stellar parameters for GJ 9827 derived by applying \textsc{SpecMatch-Emp} (described below) to data from the FIbre-fed \'Echelle Spectrograph, and \citet{rodriguez2017} reported derived stellar parameters from a principal component analysis of data from the High Accuracy Radial Velocity Searcher (HARPS) by \cite{houdebine2016}. As an independent check on the stellar parameters of GJ 9827, we compared our PFS ``template'' spectrum to a library of high-resolution ($\sim$55,000), high S/N ($>100$) Keck/HIRES stellar spectra from the California Planet Search. The stars included in the library have well-determined parameters from a variety of sources, e.g., interferometry, optical and NIR photometry, asteroseismology, and LTE analysis of high resolution optical spectra. \cite{yee2017} graciously provide this library and matching algorithm packaged in the Python codebase Empirical SpecMatch, or \textsc{SpecMatch-Emp}.\footnote{We followed the exact procedure outlined in the \textsc{SpecMatch-Emp} ``quickstart'' guide found at \url{http://specmatch-emp.readthedocs.io/en/latest/quickstart.html}.} Focusing on the spectral region around the Mgb triplet ($\sim5165-5200$~{{\AA}}), we used \textsc{SpecMatch-Emp} to shift our continuum-normalized PFS spectrum of GJ 9827 to the internal library wavelength scale, and then compare our spectrum to every other library star to find the best matches using $\chi^2$ minimization. \textsc{SpecMatch-Emp} also includes a spline continuum fit and floating $v~sin~i$ term (rotational broadening kernel) in the minimization. 

\begin{figure}
	\centering
    \includegraphics[width=1\textwidth]{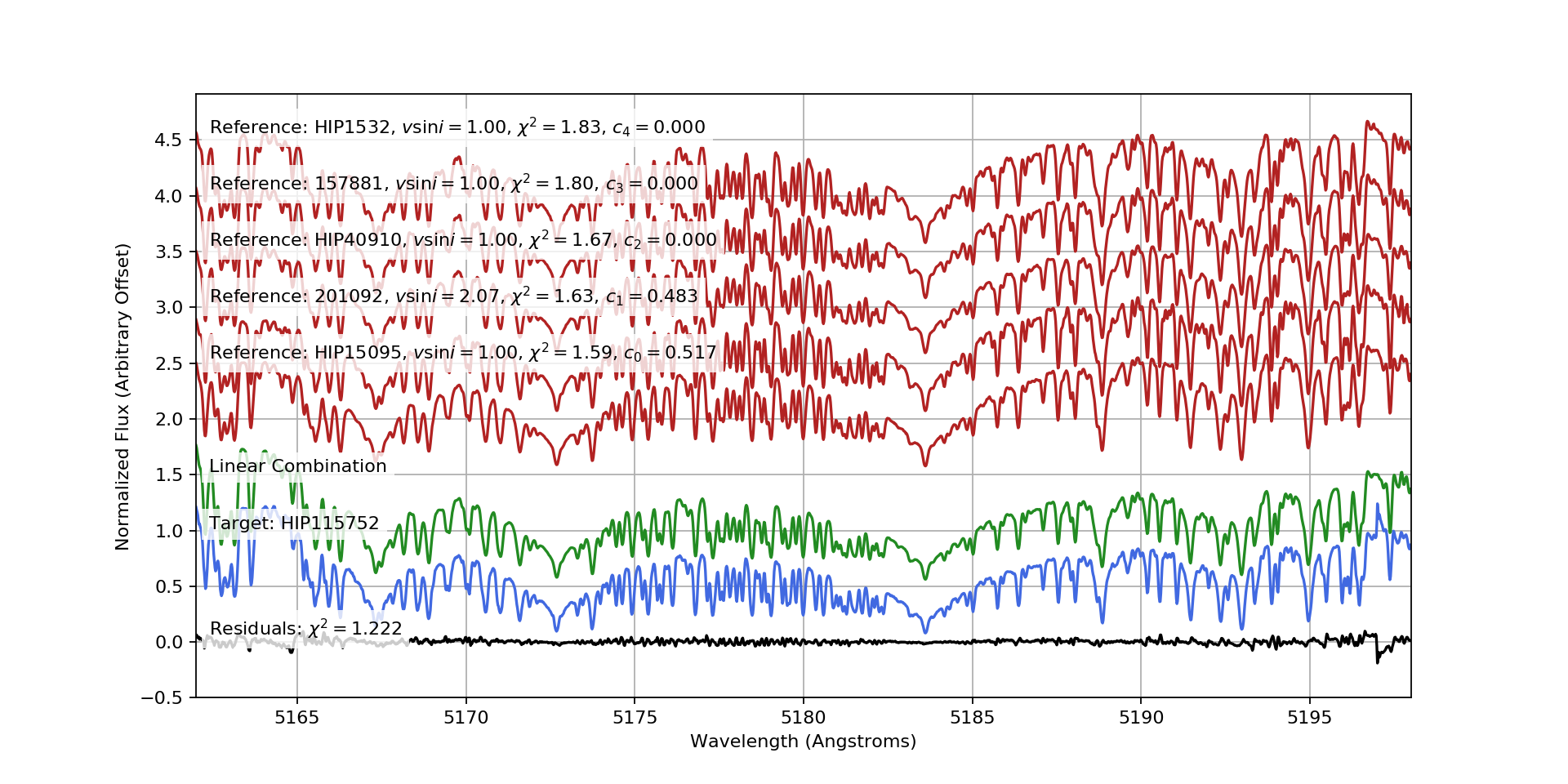}
      \label{fig:specmatch}
   \caption{\textsc{SpecMatch-Emp} library stellar spectra (red, top five spectra) that most closely match the shifted, normalized PFS ``template'' spectrum of GJ 9827 (HIP 115752, blue, bottom spectrum), and the ``best fit'' linear combination of HIP 15095 with weight 0.517 and HD 201092 with weight 0.483 (green, second from bottom spectrum). Also at the very bottom of the figure in black is the residual spectrum between the ``best fit'' and observed spectra.}
\end{figure}

From the best reference star matches, \textsc{SpecMatch-Emp} then linearly combines these reference stars, giving different weights to each reference star to achieve the closest match to the observation. In our case, only two reference stars were used in combination for the best fit parameters for GJ 9827, HIP 15095 (with weight 0.517) and HD 201092 (with weight 0.483). The matching library spectra from \textsc{SpecMatch-Emp}, the best linear combination, the PFS spectrum of GJ 9827, and the resulting residuals between the latter two are all shown in \hyperref[fig:specmatch]{Figure 1}. The best fit stellar parameters are reported in \hyperref[tab:specmatch]{Table 2}, as are the parameters derived in \citet{niraula2017} and \citet{rodriguez2017} for comparison.

\begin{table}[h]
\centering 
 \caption{GJ 9827 Stellar Parameters}
\vspace{12pt}
\begin{threeparttable}
\begin{tabular}{|c ||c | c | c||} 
\hline
Parameter & Niraula et al. (2017) & Rodriguez et al. (2017) & \textsc{SpecMatch-Emp} (this work) \\ 
 \hline
 \teff (K) & 4255$\pm$110& 4270$\pm$100  & 4085$\pm$70  \\ 
 log $g$ [cgs] & 4.70$\pm$0.15& 4.9$\pm$0.20  &4.67$\pm$0.12  \\
 $\rm{[Fe/H]}$ (dex) &-0.28$\pm$0.12 & -0.50$\pm$0.10 &-0.33$\pm$0.09  \\
 $R_{\rm{star}}$ ($R_{\odot}$) & 0.651$\pm$0.065& 0.630$^{+0.082}_{-0.077}$ &0.63$\pm$0.1900  \\
 $M_{\rm{star}}$ ($M _{\odot}$) & 0.659$\pm$0.060&0.667$\pm$0.023$^{a}$&0.62$\pm$0.08  \\
% Age (Gyr) & 9.84 $\pm$ 0.17 \\
 \hline
\end{tabular}
\begin{tablenotes}
\item[a] Stellar mass adopted in this study.
\end{tablenotes}
\end{threeparttable}
\label{tab:specmatch}
\end{table}

We find good agreement between our \textsc{SpecMatch-Emp} derived stellar parameters from the PFS template spectrum and those derived by \citet{niraula2017} and by \cite{houdebine2016} as reported in \cite{rodriguez2017}. Our \teff and log $g$ values for GJ 9827 are slightly lower (4085$\pm$70 K versus their 4255$\pm$110 K and 4270$\pm$100 K, 4.67$\pm$0.12 cgs versus their 4.70$\pm$0.15 cgs and 4.9$\pm$0.2 cgs), and our metallicity is almost the same as \citeauthor{niraula2017} (-0.30$\pm$0.09 dex versus their -0.28$\pm$0.12 dex) while slightly higher than \citeauthor{rodriguez2017} (-0.5$\pm$0.1 dex). Our \textsc{SpecMatch-Emp} derived stellar radius of 0.63$\pm$0.19 $R_{\odot}$ is slightly smaller than \citeauthor{niraula2017}'s (0.651$\pm$0.065 $R _{\odot}$), and has a larger error but is identical to \citeauthor{rodriguez2017}'s best-fit stellar radius from the transit light curves (0.630$^{+0.082}_{-0.077}$ $R _{\odot}$), and our derived stellar mass of 0.62$\pm$0.08 $M _{\odot}$ is lower than both the value reported by \citeauthor{niraula2017} and the value \citeauthor{rodriguez2017} derives from the semi-empirical mass-absolute magnitude relation from \citet{mann2015}, 0.667$\pm$0.023 $M _{\odot}$. Due to the greater precision of the \citeauthor{mann2015} stellar mass, we adopt it in our calculations of planetary mass reported below. 

Neither \citet{niraula2017} nor \citet{rodriguez2017} report an age estimate for GJ 9827. From the UVW space velocities and the relations in \cite{reddy2006}, the star has a 97\% probability of being a thin disk member, which in combination with its %\textsc{SpecMatch-Emp} derived age (9.84$\pm$0.17 Gyr) and 
low metallicity suggests the star is an older thin disk member, perhaps enhanced in $\alpha$ elements (see, for example, \citealt{haywood2013}). We leave a detailed abundance analysis of this star for future work.

We do not see a strong correlation between the PFS RV observations of GJ 9827 and either of the following two spectroscopic activity indicators: the emission flux measured in the Ca \textsc{II} H \& K lines, $S_{\rm HK}$, and the emission in the H$\alpha$ line, ``$S_{\rm H\alpha}$'' (\hyperref[fig:specmatch]{Figure 2}). $S_{\rm HK}$ is defined as in \cite{duncan1991} with the updated $R$ continuum area center from \cite{santos2000}, and $S_{\rm H\alpha}$ is defined as in \cite{gomesdasilva2011}. We performed Pearson and Kendall correlation tests on $S_{\rm HK}$ and $S_{\rm H\alpha}$ versus the RVs, and the results of the tests do not show evidence of any correlation between these quantities (the $p$-values are $>0.1$ in all cases). We also do not see evidence of any correlation between $S_{\rm HK}$ or $S_{\rm H\alpha}$ and the RV residuals against our best Keplerian fits. We also checked for significant power in the periodograms of the $S_{\rm HK}$ values and $S_{\rm H\alpha}$ and found no significant power at any period.

GJ 9827 does not appear to be a very active star based on $S_{\rm HK}$ or $S_{\rm H\alpha}$ compared to other stars with similar $B-V$ colors (its $B-V$ is 1.319; \citealt{rodriguez2017}). Compared to stars within $1.269 < B-V < 1.369$ observed in the PFS survey, GJ 9827's mean $S_{\rm HK}$ value is ranked at 24\% (meaning the value is higher than 24\% of all stars in this color bin) while the RMS of its $S_{\rm HK}$ value is ranked at 57\%. Its rankings for the value and the RMS of $S_{\rm H\alpha}$ are 17\% and 42\%, respectively. Given the lack of correlation between activity indices and RVs, we do not make an attempt to correct for any stellar activity trends in our subsequent analysis of the RV observations.

\begin{figure}
	\centering
    % this figure is made by Sharon - contact her for revision
    \includegraphics[width=1\textwidth]{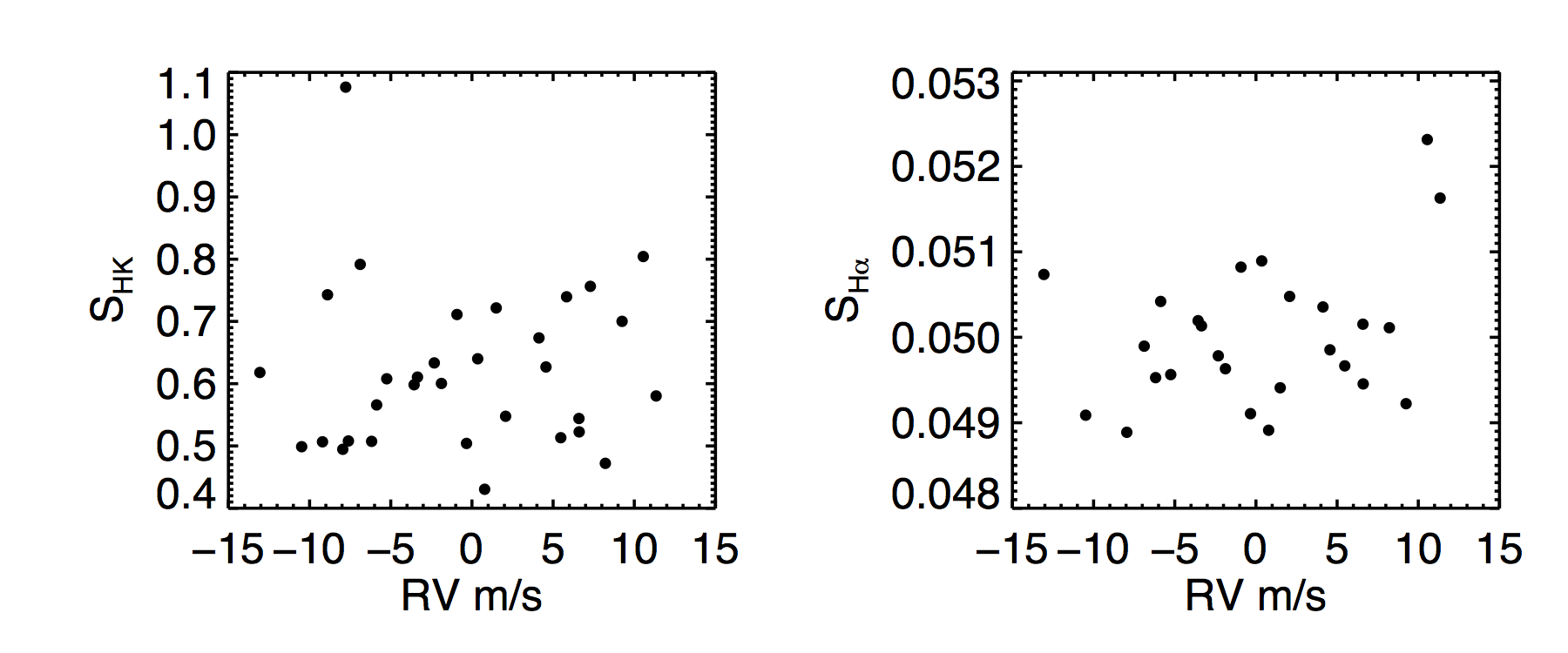}
      \label{fig:activity}
\vspace{-30pt}
\caption{Emission fluxes in the Ca \textsc{II} H \& K lines (on the right) or the H$\alpha$ line (left) plotted against the radial velocities. Extreme outliers in $S_{\rm HK}$ or $S_{\rm H\alpha}$ are not plotted, because they are likely due to errors in our measurements for the line flux. See Section~\ref{sec:stellar} for more.}
\end{figure}

\subsection{Best-Fitting Keplerian Solutions}

\subsubsection{SYSTEMIC Solution}
We used v2.173 of the SYSTEMIC console \citep{systemic} to fit Keplerian orbital models to the PFS RV data of GJ 9827. We began with a three-planet fit, holding the planetary period values constant at the measured values from \citet{rodriguez2017}, which have errors of order $\leq 6\times10^{-5}$ days (consistent with \citealt{niraula2017}; moving forward we chose to produce RV fits to the planets as reported in \citeauthor{rodriguez2017}). We also assumed circular orbits \citep[e.g.,][]{vaneylen2015}, 90$^{\circ}$ inclination orbits, although our results do not change if we adopt the inclinations reported in \citeauthor{rodriguez2017}; given the stellar radius and observed period of planet d, the inclination must be $\geq 87^{\circ}$. To ensure that the radial velocity measurements were properly phased such that at the transit mid-point the measured RV would be zero, we (1) shifted the measured time of ascending node reported by \citet{rodriguez2017} from the reference epoch of 2454833 to the epoch of our first observation, 2455198.54192, (2) determined the orbital phase of the planet at that time, and (3) converted that fractional phase into degrees. These values in degrees were then input as the planetary mean anomalies and held constant during the fit optimization process. Incorporating errors in the planetary periods result in only minor changes in the derived mean anomaly values, ranging from $\sim0.4$ to $\sim1.45$ deg between our first RV observation and the reference epoch in \citeauthor{rodriguez2017}.  

Each radial velocity measurement has an associated formal (observing error) uncertainty that varies for every observation, and an error term accounting for scatter about the fit (e.g., from underestimated measurement errors, stellar jitter, other astrophysical sources of variation) that is the same for each observation in the data set. In SYSTEMIC the best-fit parameters are then derived by optimizing the log-likelihood of the model:
\begin{equation}
\rm{log}~\mathcal{L} = -\frac{1}{2} \bigg[\chi^{2} +\sum_{{\it i=1}}^{{\it N_o}} \rm{log}~(\textit{e}^{2}_{\it i} +\textit{s}^{2}_{\it jit})+N_{o} \rm{log}~(2\pi) \bigg]
\end{equation} 

\noindent where
\begin{equation}
\chi^2 = \sum_{i=1}^{N_o} (V_i-v_i)^2/(e^2_i+s_{jit}^2),
\end{equation}

\noindent and $V_i$ is the predicted velocity, $v_i$ is the observed velocity,
$e_i$ is the formal error, and $s_{jit}$ is an additional error term, held constant across all observations. 

In a semi-automatic way, SYSTEMIC chooses the best (maximum log-likelihood) parameters for a planet fit using a downhill simplex algorithm \citep[AMOEBA, ][]{press1992} Nelder \& Mead 1965. In the fitting of the PFS GJ 9827 RV data, most of the available parameters are kept fixed, including planet period, mean anomaly (as described above), eccentricity, longitude of pericenter ($\varpi$), and inclination, allowing only the RV semi-amplitude $K$ (and the resulting planet mass) to be free parameters.\footnote{We assume the stellar mass of GJ 9827 is 0.667$\pm$0.023 $M_{\odot}$. Folding in the stellar mass error into the errors on derived planet mass increases these values by $\leq$0.01 $M_{\oplus}$.} In \hyperref[tab:fit]{Table 3} we report all of the fixed parameters and the best-fit, median, and median absolute deviation (MAD, a measure of the dispersion in a set of data that is less sensitive to outliers than the standard deviation) values for the fitted planet parameters, planet mass and corresponding radial velocity semi-amplitude.  

The median and MAD values for the planet parameters come from refitting the data to derive new parameters via a Markov-Chain Monte Carlo algorithm \citep[MCMC, e.g.][]{ford2005,ford2006,gregory2011} within SYSTEMIC, paired with a flat prior on log $M$ with a range between 10$^{-4}$ M$_J$ and 1000 M$_J$. %paired with flat priors on $K$. 
As noted in \hyperref[tab:fit]{Table 3}, the rest of the parameters were fixed. We ran the MCMC in Systemic until the chains were well-mixed, defined in SYSTEMIC as all parameters having a Gelman-Rubin statistic of $\leq 1.005$ \citep{gelman2003,ford2006}. The values reported in \hyperref[tab:fit]{Table 3} come from an MCMC with 10 chains, skipping the first 1000 iterations as a ``burn-in'' period, and only retaining one model in every 170 to minimize correlations between nearby elements of the Markov chain; the resulting total chain length is 1.67$\times 10^{5}$. We also tried increasing the number of iterations that are skipped and discarded, and continued to achieve these same posterior distribution.

We also included as free parameters a ``jitter'' term, $s_{jit}$. In \citet{teske2016} we outlined an analytic approach for estimating $s_{jit}$ that can be implemented outside of SYSTEMIC and without including the term as a free parameter, and showed that it produced results consistent with the MCMC distributions. Thus here, as in previous works using SYSTEMIC \citep[e.g.,][]{rowan2016,vogt2015}, we determine $s_{jit}$ via MCMC using a modified Jeffrey function as a prior \citep[see ][]{vogt2015}. Including a linear trend term in the fit resulted in slightly poorer RMS and $-\rm{log}~\mathcal{L}$, and did not improve the constraints on $K$, so we left out this term in our final results. 

\begin{figure*}[t]
\centering
\includegraphics[width=0.4\textwidth]{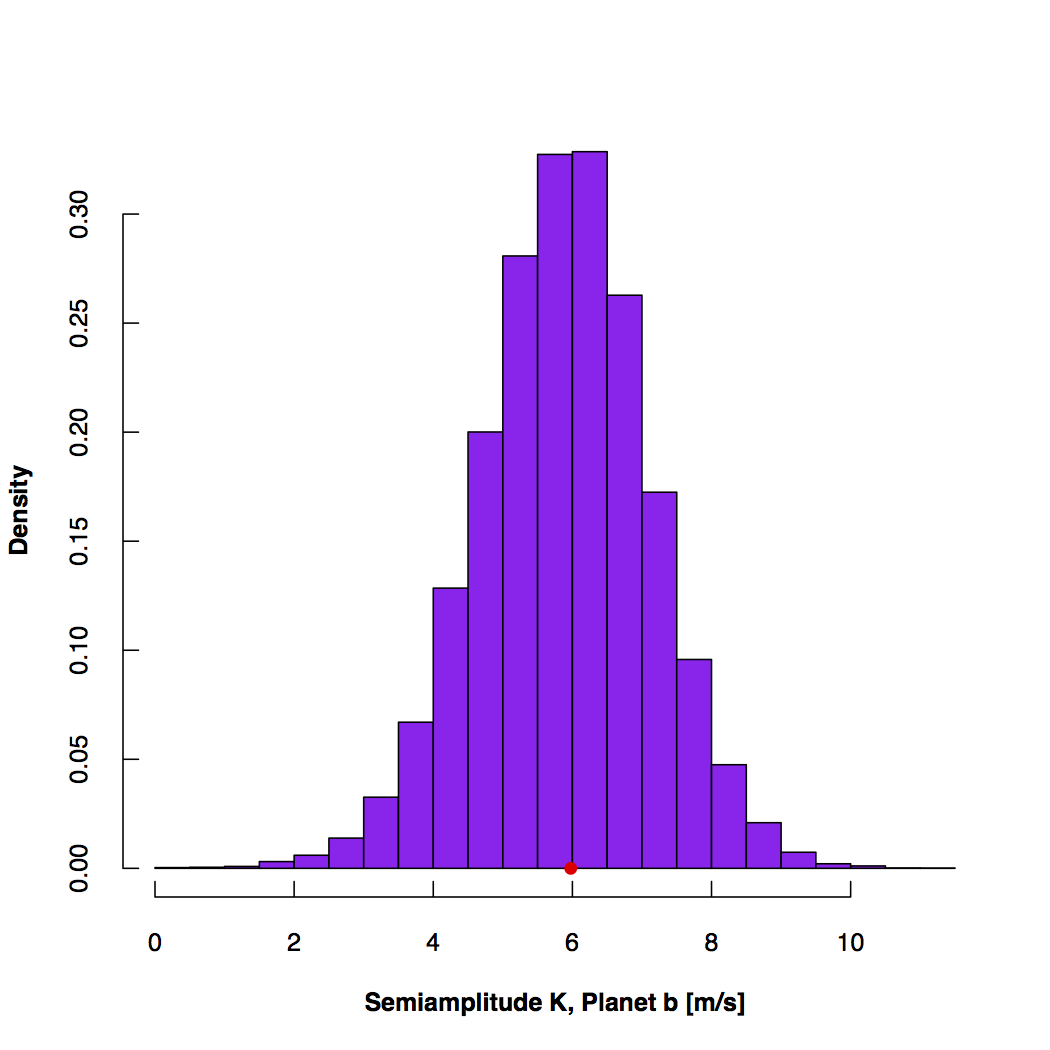}
\includegraphics[width=0.4\textwidth]{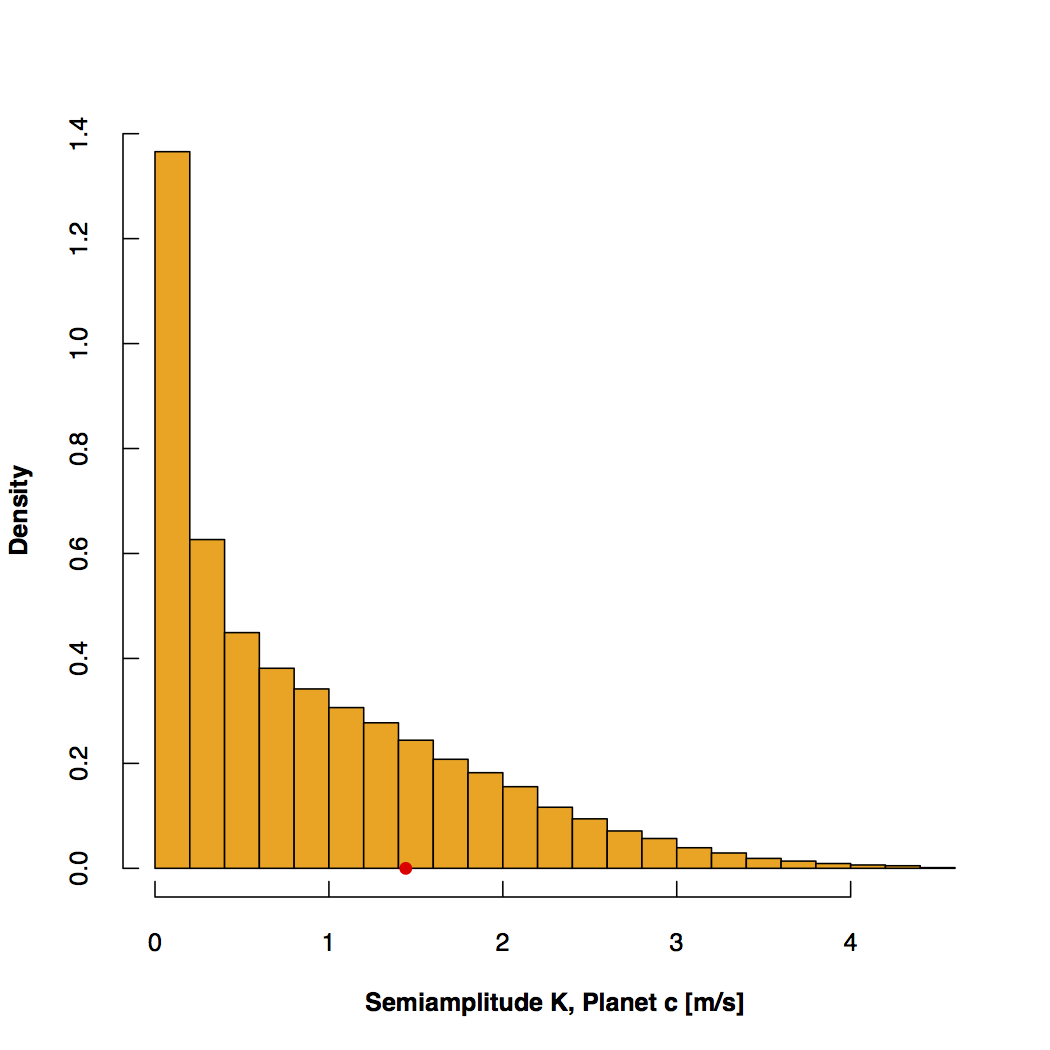}
\includegraphics[width=0.4\textwidth]{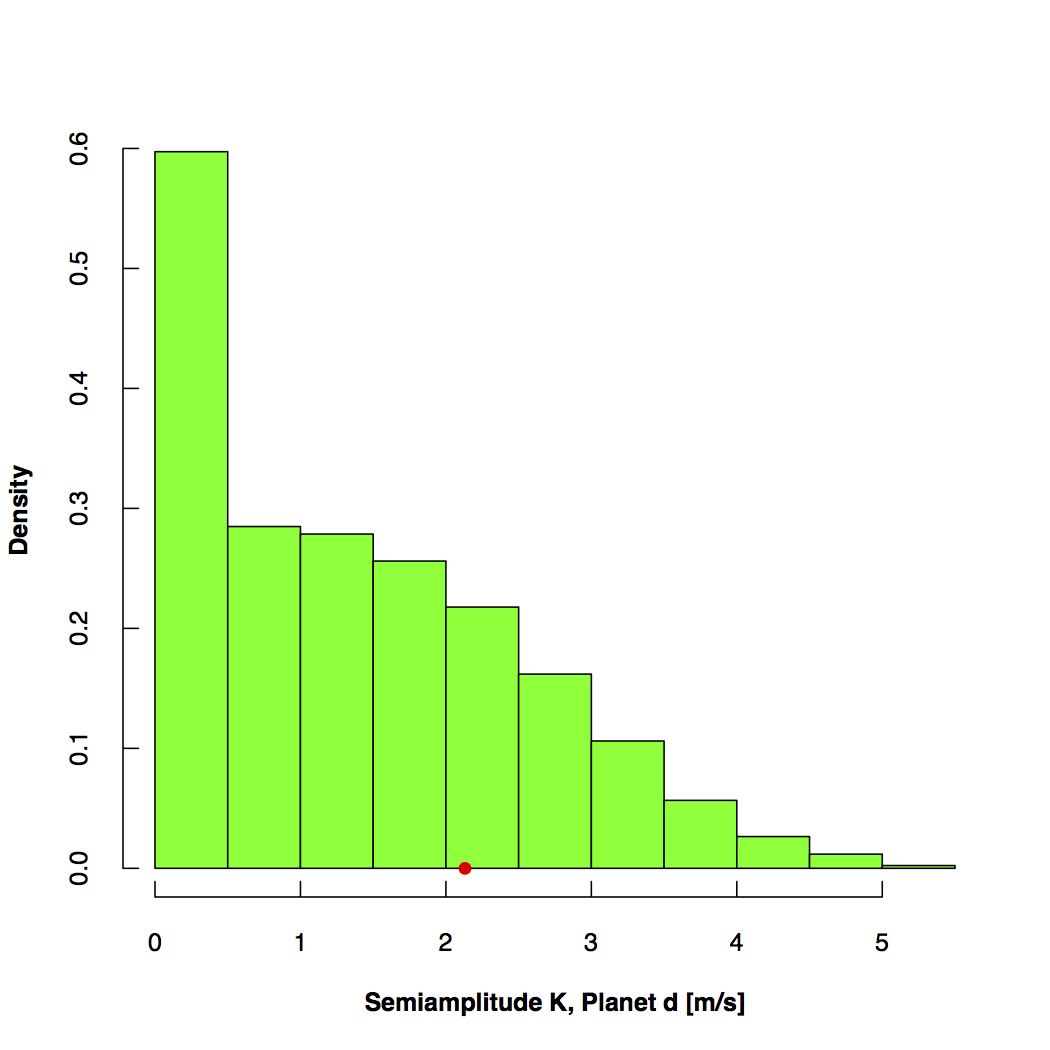}
\includegraphics[width=0.4\textwidth]{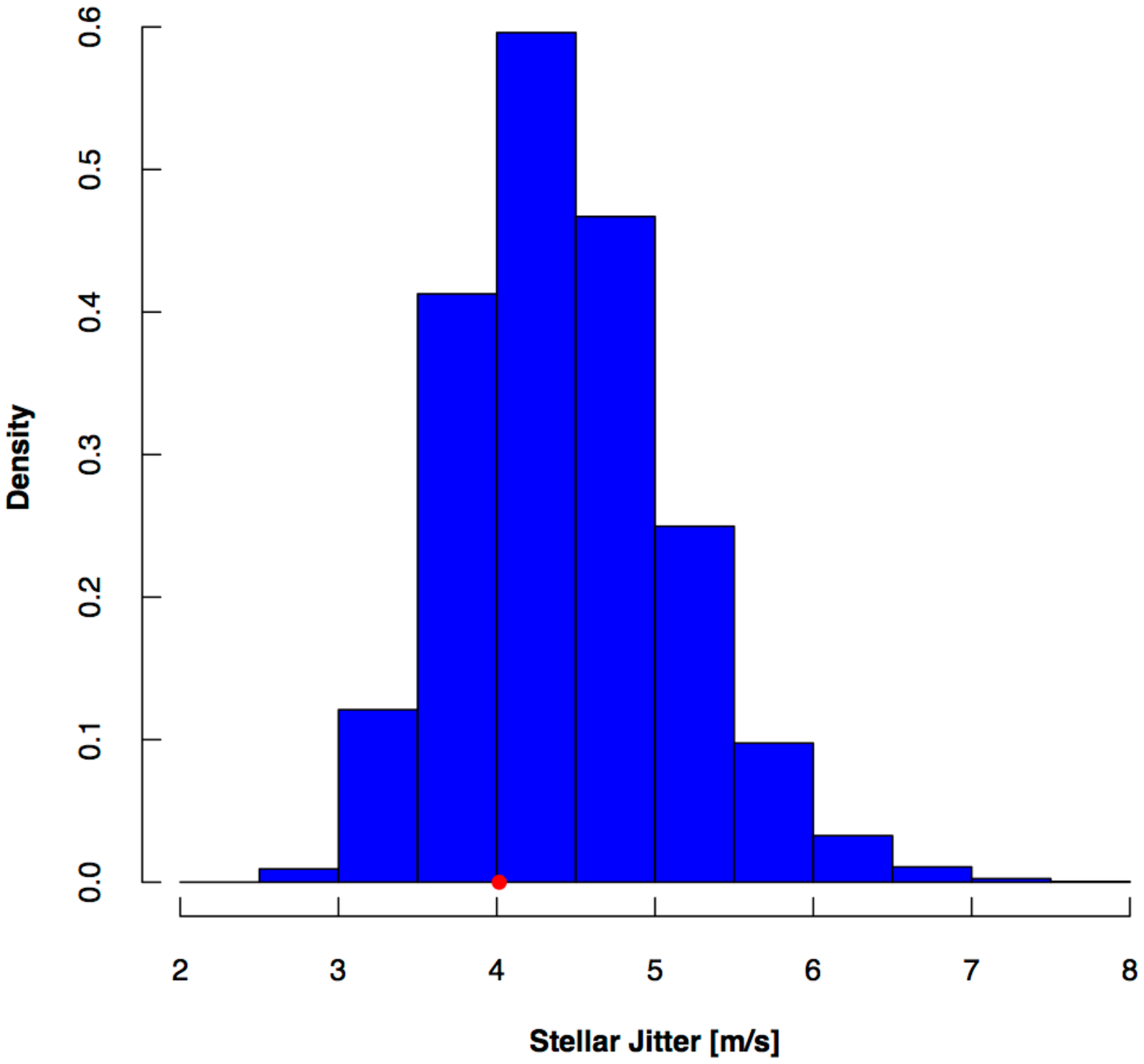}
\caption{Marginalized posterior distributions of the planet RV semiamplitudes (b in purple, c in orange, and d in green) and jitter (in blue) resulting from our MCMC analysis of GJ 9827 PFS data using SYSTEMIC. The best-fit values from \hyperref[tab:fit]{Table 3} are marked with red dots.} \label{fig:systemic_mcmcerrors}
\end{figure*}

As shown in \hyperref[fig:systemic_mcmcerrors]{Figure 3}, the MCMC posterior distributions for the outer planet semi-amplitudes $K_c$ and $K_d$ are peaked at zero, and instead indicate that the PFS data are only able to place upper limits on the masses of these planets. That said, we also performed an analysis using \textsc{RadVel} (\S3.2.2), which shows $K$ posterior distributions peaked away from zero; the difference in the mode of the posterior distributions is likely due to a difference in the priors: SYSTEMIC uses a uniform prior on log $M$, which transforms to a skewed prior on $M$ with a mode at zero, while our implementation of \textsc{Radvel} used a uniform prior on $K$, which transforms to a uniform prior on $M$.
%could be due to differences between the MCMC algorithms used by the two software packages (a generic Metropolis sampler in SYSTEMIC vs. the emcee \citep{foreman-mackey2013} affine invariant ensemble sampler in \textsc{RadVel}). 
%However, based on an analysis using \textsc{RadVel} presented in \S3.2.2 that shows  $K$ posterior distributions not peaked at zero, these upper limits may be an anomalous result of the MCMC algorithm used by SYSTEMIC, perhaps getting caught in a local minimum.  AW: but the chains are well-mixed (as indicated by a low final R-hat value), so getting caught in a local minimum is unlikely to be the problem.  Not impossible, but unlikely

\begin{figure*}[t]
\centering
\includegraphics[width=0.4\textwidth]{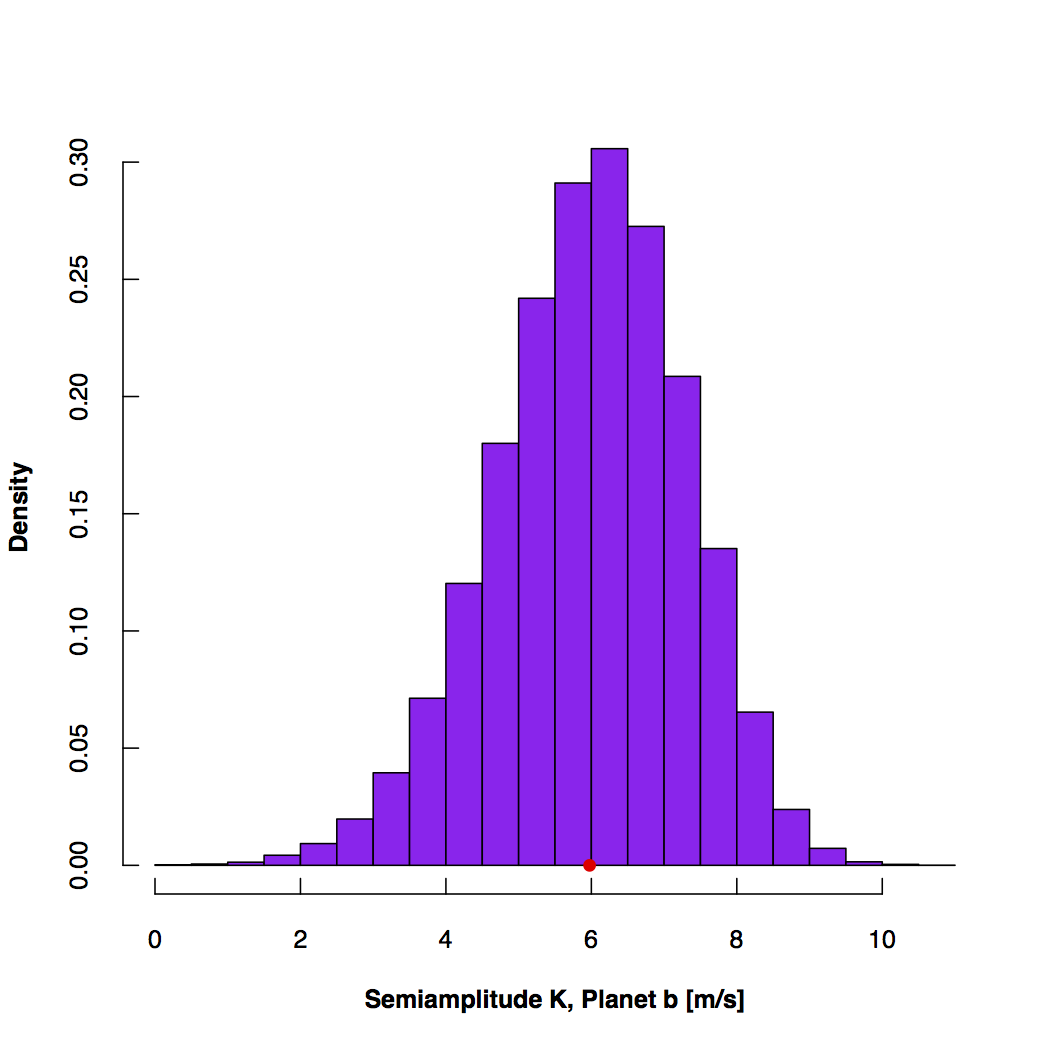}
\includegraphics[width=0.4\textwidth]{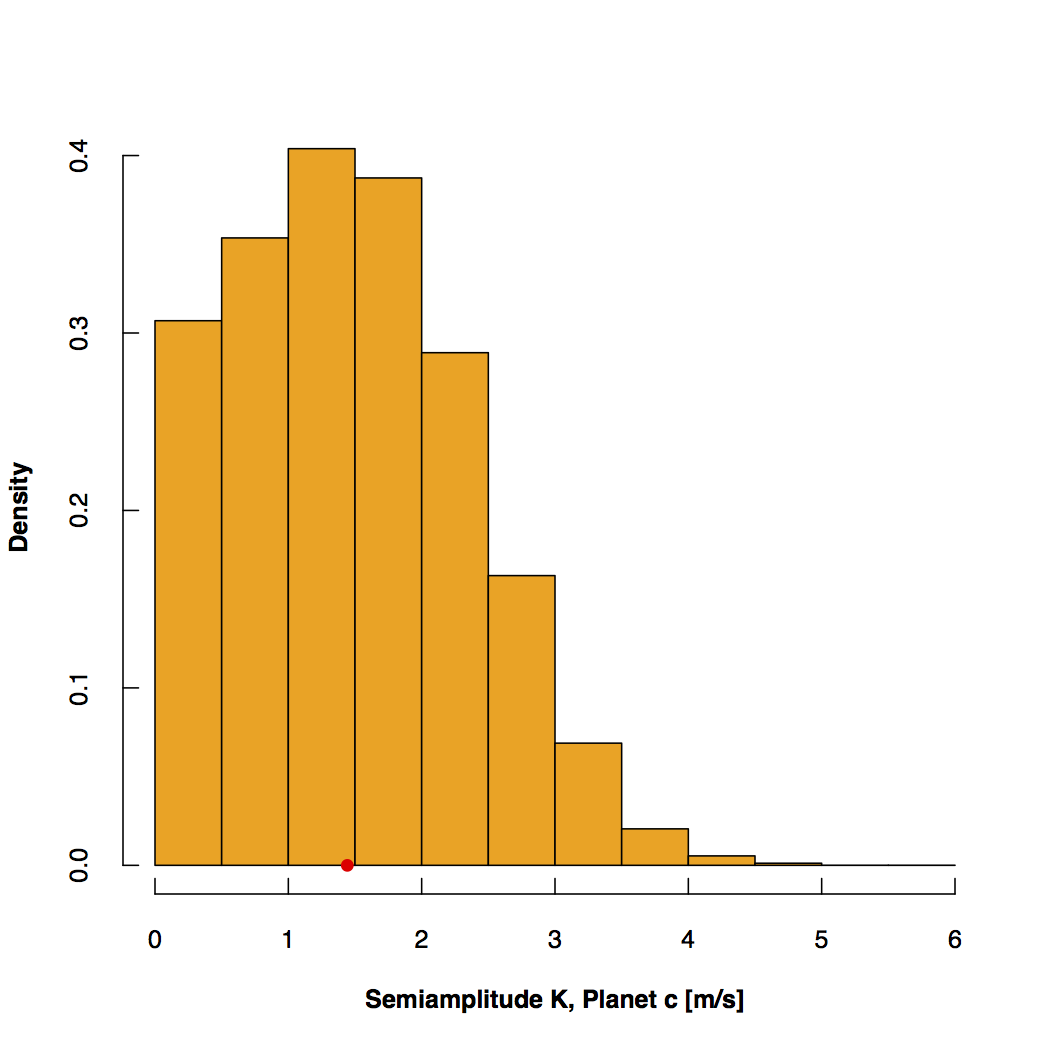}
\includegraphics[width=0.4\textwidth]{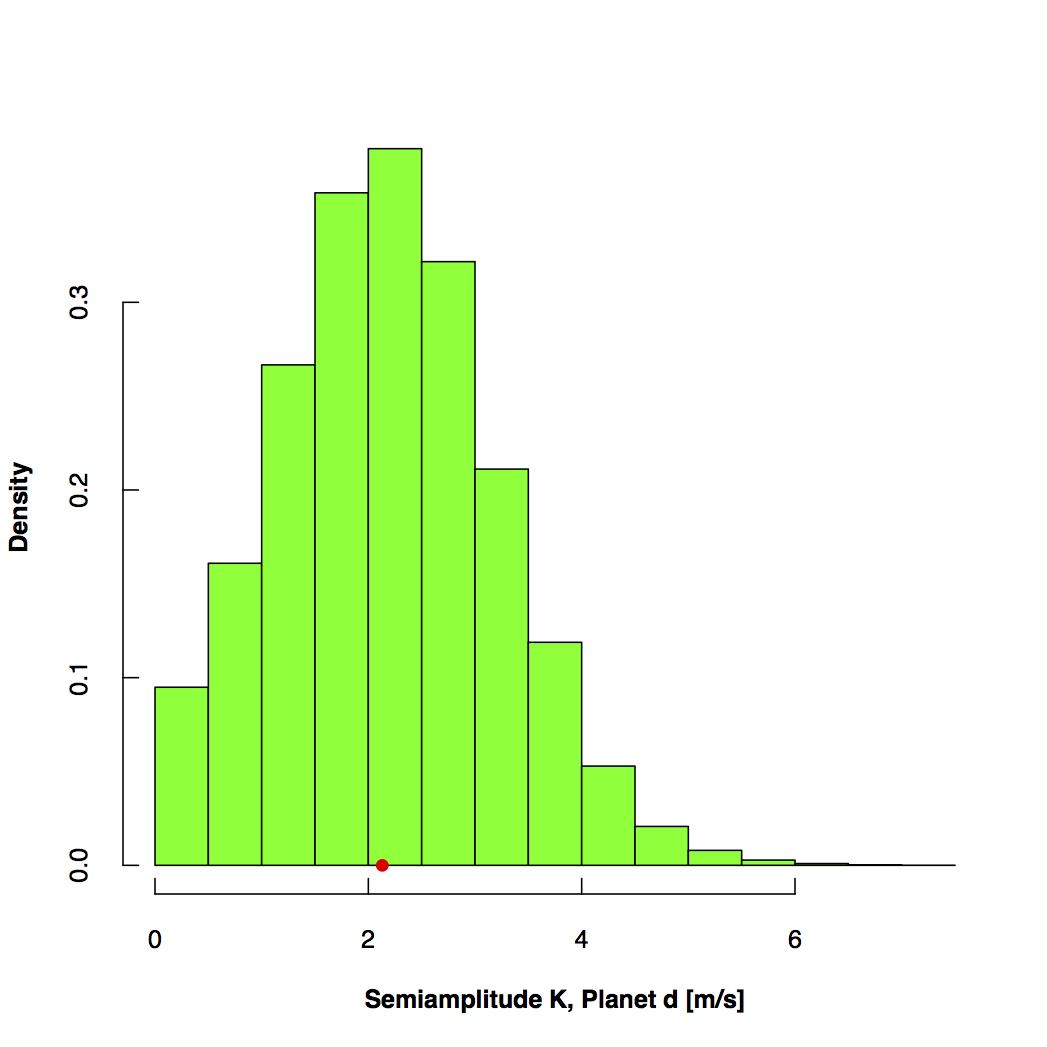}
\includegraphics[width=0.4\textwidth]{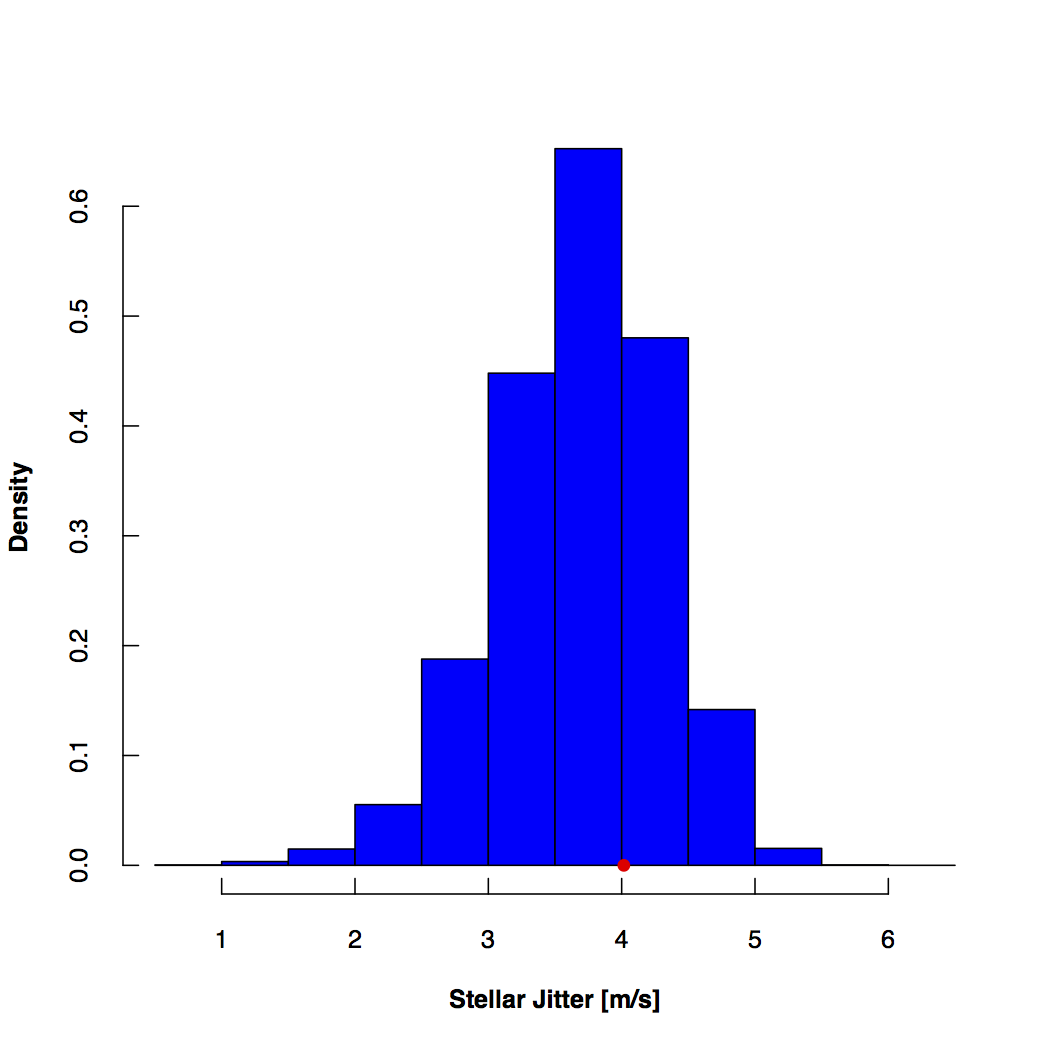}
\caption{Marginalized posterior distributions of the planet RV semiamplitudes (b in purple, c in orange, and d in green) and jitter (in blue) resulting from our bootstrap error analysis of GJ 9827 PFS data using SYSTEMIC. The best-fit values from \hyperref[tab:fit]{Table 3} are marked with red dots.} \label{fig:systemic_bootstraperrors}
\end{figure*}

For comparison, we also present in \hyperref[tab:fit]{Table 3} the errors resulting from a bootstrap error estimation \citep{press1992} done in SYSTEMIC, using 100,000 trails. The distributions of bootstrapped errors on $K$ (\hyperref[fig:systemic_bootstraperrors]{Figure 4}) are much more similar to the results of the \textsc{RadVel} analysis. However, we note that the implementation of bootstrapping errors in SYSTEMIC are incorrectly drawn in sample space instead of in likelihood space, as described in \citealt{loredo2013}, and therefore \textit{should not be trusted}. Further discussion of the differences between error estimates is in \S4.2. 

%Neither SYSTEMIC nor \textsc{RadVel} 
SYSTEMIC does not take into account the errors on the stellar mass in the calculation of planet mass from the RV semi-amplitude. Due to the small error on the mass of GJ 9827 (0.023 M$_{\odot}$, \citealt{rodriguez2017}), propagating the stellar mass error in the planet mass calculation results in almost no change ($\sim$1\% increase) in the planet mass errors. For ease of reproducibility of our results, we thus do not include the stellar mass errors in the planet mass errors reported here %for either SYSTEMIC or \textsc{RadVel}.
for SYSTEMIC.

\begin{longrotatetable}
%\startlongtable
\begin{deluxetable}{cccccccccc}
\tablecaption{Kepler Fit Solution from SYSTEMIC for planets around GJ 9827\label{tab:fit}}
\tablecolumns{10}
\tablewidth{0pt}
\tabletypesize{\scriptsize}
\tablehead{ 
\colhead{Parameter} & \multicolumn{3}{c}{b} & \multicolumn{3}{c}{c} & \multicolumn{3}{c}{d}\\
\cmidrule(lr){2-4} \cmidrule(lr){5-7} \cmidrule(lr){8-10}
\colhead{} & \colhead{Best-fit} & \colhead{Median$\pm$MAD} & \colhead{Median$\pm$MAD}  & \colhead{Best-fit} & \colhead{Median$\pm$MAD} & \colhead{Median$\pm$MAD}  & \colhead{Best-fit} & \colhead{Median$\pm$MAD} & \colhead{Median$\pm$MAD}  \\ 
\colhead{} & \colhead{} & \colhead{MCMC} & \colhead{Bootstrap} & \colhead{} & \colhead{MCMC} & \colhead{Bootstrap} & \colhead{} & \colhead{MCMC} & \colhead{Bootstrap}
}
\startdata
Period\tablenotemark{a} [days] & 1.2089819 [fixed] & \nodata &  \nodata & 3.648086 [fixed]  & \nodata &  \nodata & 6.201472 [fixed] & \nodata &  \nodata \\ 
$M$ [$M_{\oplus}$] & 7.60 & 7.50$\pm$1.52 & 7.67$\pm$1.66 & 2.65 & $<$2.56 %1.14$\pm$1.42 
& 2.60$\pm$1.75 & 4.67 & $<$5.58 %2.66$\pm$2.92 
& 4.73$\pm$2.26 \\ 
Mean anomaly\tablenotemark{b} [deg] & 26.5974900 [fixed] & \nodata & \nodata & 357.0871081 [fixed] & \nodata &  \nodata & 100.7029299 [fixed] & \nodata & \nodata \\ 
Eccentricity  & 0 [fixed] & \nodata & \nodata & 0 [fixed] & \nodata  & \nodata & 0 [fixed] & \nodata &  \nodata \\ 
$\varpi$ [deg] & 0 [fixed] & \nodata & \nodata & 0 [fixed] & \nodata  & \nodata & 0 [fixed] & \nodata &  \nodata  \\ 
$K$ [m\,s$^{-1}$] & 5.98 & 5.90$\pm$1.20 & 6.04$\pm$1.31 & 1.44 & $<$1.39 %0.62$\pm$0.77 
& 1.42$\pm$0.96 & 2.13& $<$2.54 %1.21$\pm$1.33 
& 2.16$\pm$1.03\\ 
Semi-major axis\tablenotemark{a} [AU] & 0.0211 [fixed] & \nodata  & \nodata & 0.0440 [fixed]  & \nodata & \nodata & 0.0627 [fixed]  & \nodata & \nodata\\ 
\hline
  Jitter [m\,s$^{-1}$] & 4.02 &4.38$\pm$0.66& 3.27$\pm$0.60 & & & & & & \\
\hline
\hline
Stellar mass\tablenotemark{a} [$M_{\odot}$] & 0.667$\pm$0.023 &  &  & & & & & &  \\ 
  Reduced $\chi^2$ & 1.94 &  &  & & & & & & \\
  $\rm{log}~\mathcal{L}$ & 72.40 &  &  & & & & & & \\
  RMS [m\,s$^{-1}$] & 4.60 &  &  & & & & & & \\ 
  Data points  & 36&  &  & & & & & & \\ 
  Span of observations [JD] & 2455198.54$-$2457615.79 &  &  & & & & & & \\
\enddata
\tablenotetext{a}{Held constant during fit optimization, from \citet{rodriguez2017}.}
\tablenotetext{b}{Held constant during fit optimization, derived from \citet{rodriguez2017}.}
\tablecomments{All elements are defined at epoch JD = 2455198.54192.}
\end{deluxetable}
\end{longrotatetable}

%NEED FIGURE HERE SHOWING RV FITS TO PFS DATA. HAVING A HARD TIME GETTING RIGHT WITH SYSTEMIC. Update: Giving up.

From our SYSTEMIC analysis, we find that GJ 9827b has a best-fit mass of 7.60 M$_{\oplus}$ and a median mass of 7.50$\pm$1.52 M$_{\oplus}$ %and 7.67$\pm$1.66 M$_{\oplus}$ 
from the MCMC %and bootstrap analysis, respectively (this order holds in the next two sentences, as well). 
analysis. We find that GJ 9827c has a best-fit mass of 2.65 M$_{\oplus}$ and a median mass $<$2.56 M$_{\oplus}$ %of 1.14$\pm$1.42 M$_{\oplus}$, 
and that %and 2.60$\pm$1.75 M$_{\oplus}$. 
planet GJ 9827d has a best-fit mass of 4.67 M$_{\oplus}$ and a median mass of $<$5.59 M$_{\oplus}$.
%2.68$\pm$2.91 M$_{\oplus}$. 
%and 4.73$\pm$2.26 M$_{\oplus}$. 
From the MCMC analysis alone, the PFS data are only able to place upper limits on the RV semi-amplitudes and masses of planets c and d. %However, our bootstrap analysis provides better, perhaps more accurate constraints when considered in tandem with the results presented in the next section.

\subsubsection{RadVel Solution}

As an independent comparison to our SYSTEMIC fitting procedure, we also used the new \textsc{RadVel} Python toolkit (v0.9.7) for modeling radial velocity data \citep{radvel}\footnote{Documentation is available at \url{http://radvel.readthedocs.io}}. \textsc{RadVel} allows the user to input multiple instrument RV data sets, specify various fit parameters (e.g., RV semiamplitude, planet period, time of inferior conjunction, eccentricity, argument of periastron; velocity zero-points and jitter terms for different instruments; linear or quadratic acceleration term), set initial guesses for those parameters and which ones to fix, and then optimize the parameters through a maximum a posteriori (MAP) fitting. %by maximizing the likelihood function, similar to SYSTEMIC above. 
 The user also has the option to add priors on any of the initialized parameters.  \textsc{RadVel} also offers an MCMC option for estimating the uncertainties on any fit parameter, and uses the Gelman-Rubin statistic ($\leq1.01$) and minimum number of independent draws ($\geq1000$) to determine if the nsembles are well-mixed. 

\begin{figure}[t]
\centering
\includegraphics[width=0.6\textwidth]{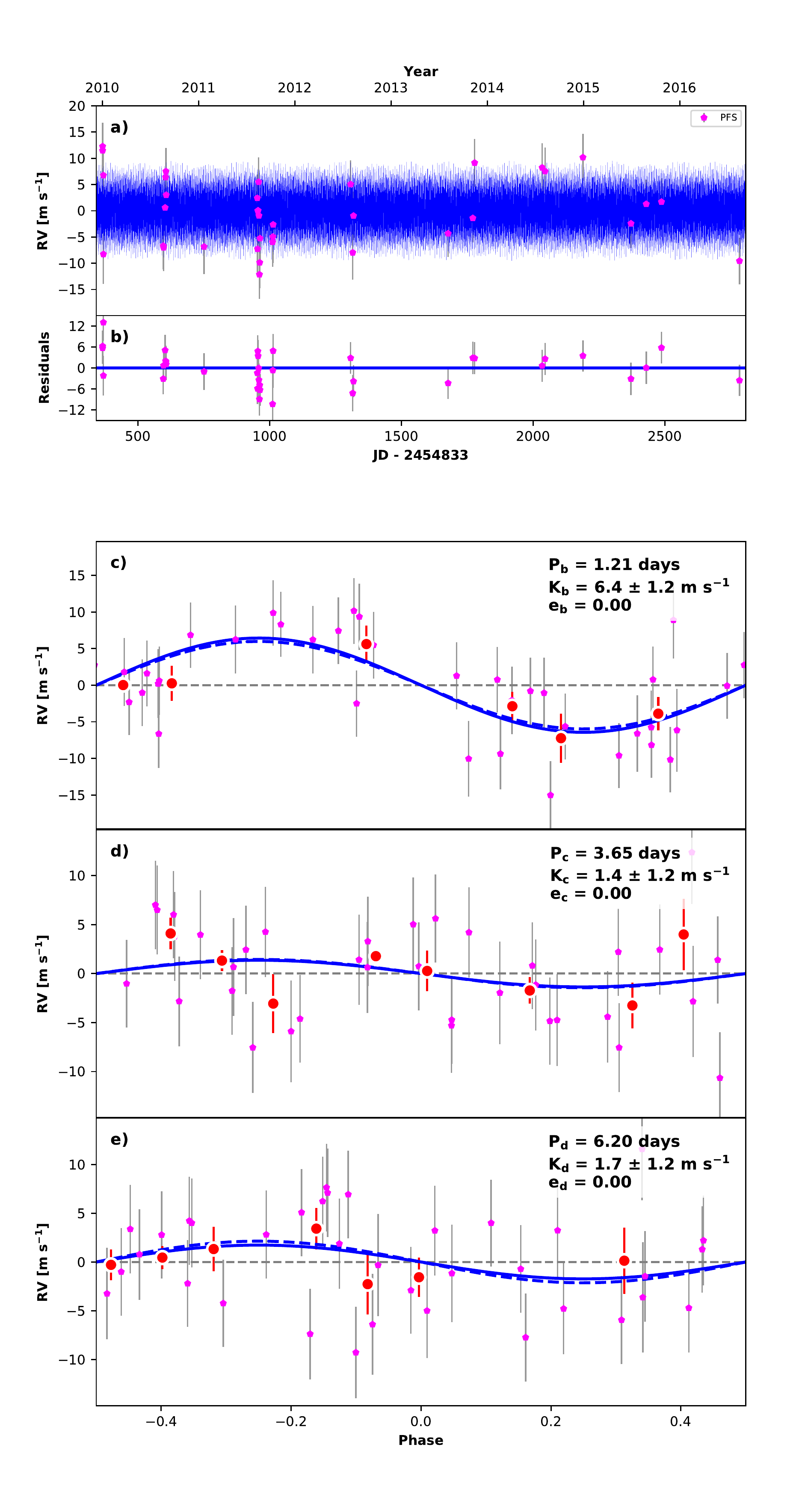}
\vspace{-30pt}
\caption{Subplot \textbf{a}: Maximum likelihood Keplerian model from \textsc{RadVel} (blue) plotted with the GJ 9827 PFS RV data (magenta). \textbf{The plotted uncertainties are the quadrature sum of the measurement uncertainties and the best-fit jitter term.} Subplot \textbf{b}: The residuals to the fit in \textbf{a}. Subplots \textbf{c}, \textbf{d}, and \textbf{e}: The phased, individual planet (best) RV fits from \textsc{RadVel}, with RVs \textbf{binned} in 0.08 units of orbital phase shown in red and the fits shown in solid blue. The upper left corner of each of these subplots also lists the planet period and eccentricity (held constant), and the maximum likelihood (best-fit) RV semi-amplitude $K$. For comparison, the curves over-plotted in dashed blue represent the SYSTEMIC best fit RV semi-amplitudes. \label{fig:radvel1}}
\end{figure}

\begin{figure}[t]
\centering
\includegraphics[width=1\textwidth]{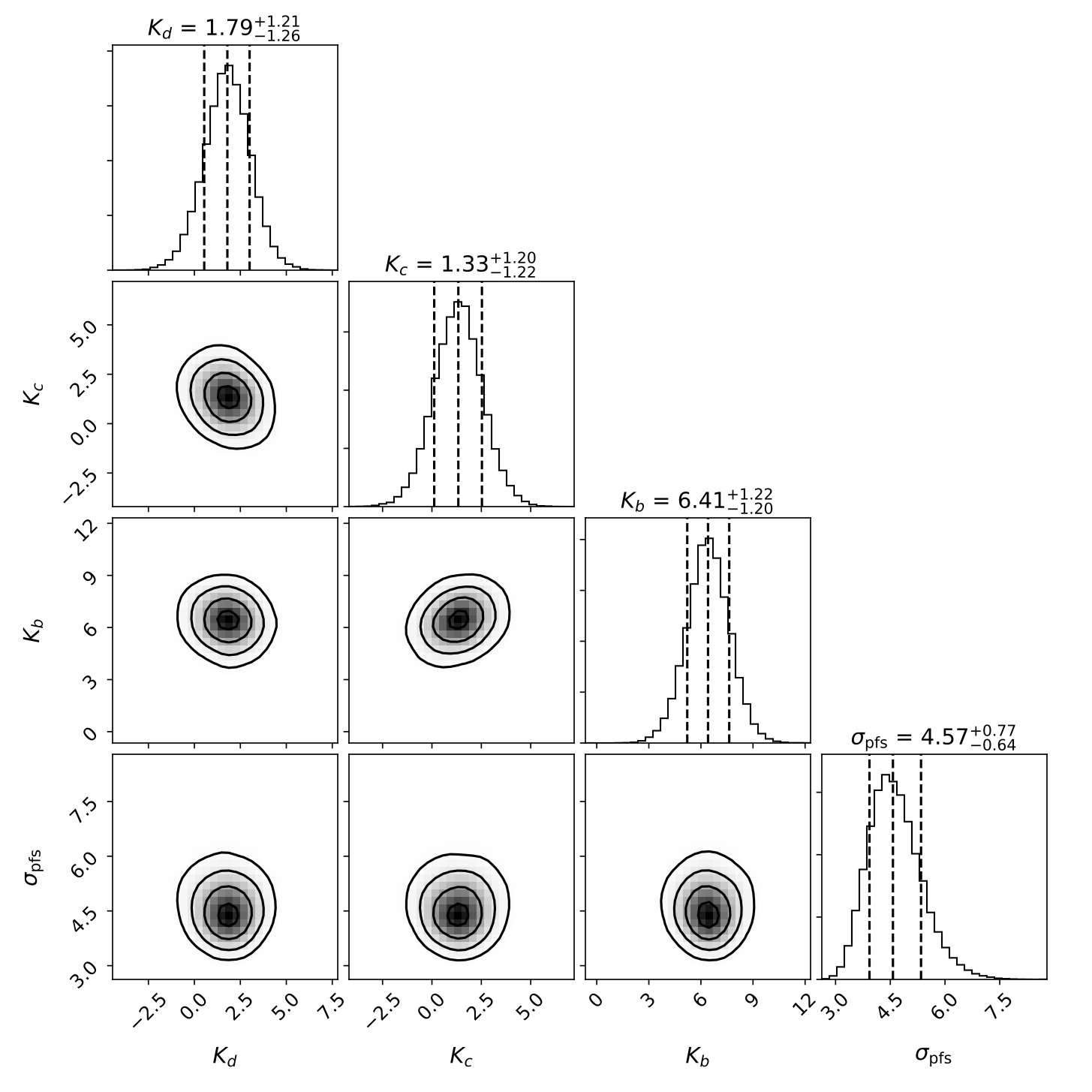}
\caption{Marginalized posterior distributions of the planet RV semi-amplitudes and  jitter resulting from our MCMC analysis of GJ 9827 PFS data using \textsc{RadVel}. \label{fig:radvel2_1}}
\end{figure}

\begin{figure}[t]
\centering
\includegraphics[width=1\textwidth]{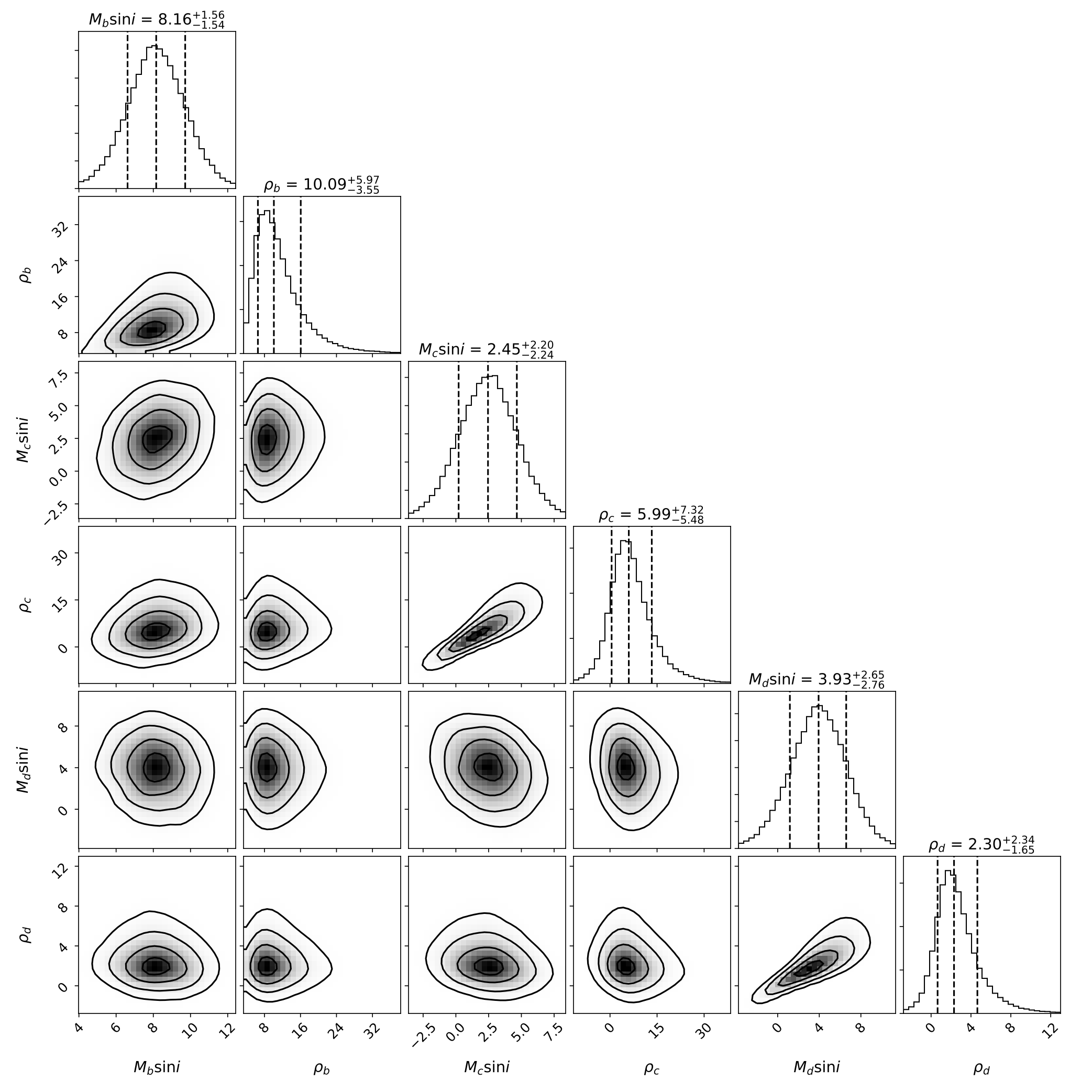}
\caption{Marginalized posterior distributions of the derived planet densities and masses (right) resulting from our MCMC analysis of GJ 9827 PFS data using \textsc{RadVel}. \label{fig:radvel2_2}}
\end{figure}

We used the same constant input values in \textsc{RadVel} as in SYSTEMIC, except instead of specifying a mean anomaly we used the mid-transit times from \citep{rodriguez2017} for the time of interior conjunction of each planet. The \textsc{RadVel} basis in this case was set to `per tc secosw sesinw k', with uniform priors ranging from -inf to inf in the basis parameters. For the MCMC fit, %we set the prior on the jitter to be between 0 and 15 m~s$^{-1}$, and set a prior on $K$ for each planet to be $>0$, and 
we ran 8 chains of 50 walkers and up to 10000 steps each. We found that the MCMC chains converged after 30000 steps. The results of the MCMC are reported in \hyperref[tab:fit2]{Table 4} and shown in Figures \hyperref[fig:radvel1]{5}, \hyperref[fig:radvel2_1]{6}, and \hyperref[fig:radvel2_2]{7}.\footnote{We provide the MCMC samples and associated relevant files here: \url{https://github.com/jkteske/GJ9827.git}.} We also tried (1) setting the prior on the jitter to be between 0 and 15 m~s$^{-1}$ and setting the prior on $K$ for each planet to be $>0$, and (2, separately) setting the prior on $K$ for each planet to be a Gaussian distribution with $\mu=3$ m~s$^{-1}$ and $\sigma=$ 30 m~s$^{-1}$; none of these changes in the prior changed our final results within the quoted errors. 

From our \textsc{RadVel} analysis, we find that GJ 9827b has best-fit mass 8.20 %$\pm$1.53 
M$_{\oplus}$  and a median mass of 8.16$^{+1.56}_{-1.54}$ M$_{\oplus}$ (where the $\pm$ represents the full 68$^{th}$ percentile credible interval), GJ 9827c has a best-fit mass of 2.50 %$\pm$2.21 
M$_{\oplus}$ and a median mass of 2.45$^{+2.20}_{-2.24}$ M$_{\oplus}$, and GJ 9827d has a best-fit mass of 3.80 %$\pm$2.63 
M$_{\oplus}$ and a median mass of 3.93$^{+2.65}_{-2.76}$ M$_{\oplus}$. As expected based on the SYSTEMIC analysis results, \textsc{RadVel} finds poorer constraints on the RV semi-amplitudes and masses of planets c and d; in both cases the PFS data are only really able to place upper limits on these planets' masses. We note that the upper limits fall within the ranges predicted by \cite{wolfgang2016} -- 3.8$\pm$1.9 M$_{\oplus}$ and 7.0$\pm$1.9 M$_{\oplus}$ -- based on the intrinsic astrophysical dispersion in planet masses at the given radius.

\startlongtable
\begin{deluxetable}{ccccccc}
\rotate 
\tablecaption{Keplerian Fit Solution from \textsc{RadVel} for planets around GJ 9827 \label{tab:fit2}}
\tablecolumns{7}
\tablewidth{0pc}
\tabletypesize{\scriptsize}
\tablehead{ \colhead{Parameter} & \multicolumn{2}{c}{b} &\multicolumn{2}{c}{c} & \multicolumn{2}{c}{d} \\
\cmidrule(lr){2-3} \cmidrule(lr){4-5} \cmidrule(lr){6-7}
\colhead{} & \colhead{Best Fit\tablenotemark{a}} & \colhead{Median$\pm$34\%} & \colhead{Best Fit} & \colhead{Median$\pm$34\%} & \colhead{Best Fit} & \colhead{Median$\pm$34\% } }
\startdata
Period\tablenotemark{b} [days] & 1.2089819 [fixed]  & \nodata & 3.648086 [fixed] &  \nodata & 6.201472 [fixed] & \nodata \\ 
  $M$ [$M_{\oplus}$] & 8.20%$\pm$1.53 
  & 8.16$^{+1.56}_{-1.54}$ & 2.50%$\pm$2.21 
  &2.45$^{+2.20}_{-2.24}$ & 3.80%$\pm$2.63 
  & 3.93$^{+2.65}_{-2.76}$  \\ 
 Time of inferior conjunction\tablenotemark{a} [BJD$_{\rm{TDB}}$] & 2905.82582 [fixed] & \nodata & 2909.19940 [fixed]& \nodata & 2907.96107 [fixed] & \nodata \\ 
  Eccentricity  & 0 [fixed] & \nodata & 0 [fixed] & \nodata & 0 [fixed]  & \nodata \\ 
 $\varpi$ [deg] & 0 [fixed] & \nodata & 0 [fixed]& \nodata & 0 [fixed] & \nodata \\ 
  $K$ [m\,s$^{-1}$] &6.44%$\pm$1.2 
  & 6.41$^{+1.22}_{-1.20}$
  & 1.36%$\pm$1.2 
  & 1.33$^{+1.20}_{-1.22}$
  & 1.73%$\pm$1.2 
  & 1.79$^{+1.21}_{-1.26}$\\ 
  Semi-major axis\tablenotemark{a} [AU] & 0.0211 [fixed] & \nodata & 0.0440 [fixed] & \nodata &0.0627 [fixed] & \nodata \\ 
\hline
  Jitter [m\,s$^{-1}$] & 4.21%$\pm0.71$ 
  & 4.57$^{+0.77}_{-0.64}$ &  &  &  &  \\
  $\rm{log}~\mathcal{L}$ & 106.98 & 108.50$^{+0.86}_{-1.55}$ &  &  &   &  \\
\hline
\hline
Stellar mass\tablenotemark{a} [$M_{\odot}$] & 0.667$\pm$0.023 &  &  &  &  &  \\ 
  RMS [m\,s$^{-1}$] & 4.81 &  &  &  &  &\\ 
 Data points  & 36&  &  & & &  \\ 
  Span of observations [JD] & 2455198.54$-$2457615.79   &  & & &  \\                                     
\enddata
\tablenotetext{a}{\textsc{RadVel} performs maximum a posteriori (MAP) optimization separately from \textsc{emcee} using the Powell method as implemented in scipy.optimize.minimize.}
\tablenotetext{b}{Held constant during fit optimization, from \citet{rodriguez2017}.}
%\tablenotetext{b}{Held constant during fit optimization, derived from \citet{rodriguez2017}.}
\tablecomments{All elements are defined at epoch JD = 2454833.}
\end{deluxetable}

We performed one more test to diagnose the influence of stellar activity in our derived RV semi-amplitudes: we added a fourth periodic signal with zero eccentricity to represent the putative rotation period of GJ 9827.  The {\it K2} light curve analysis of \cite{rodriguez2017} suggests that this period is either 31$\pm$1 days or 16$\pm$1 days; we tested both. The $K$ values for the fourth periodic signal that minimize the log-likelihood in \textsc{RadVel} range from $\sim$1 to $\sim$2.4 m~s$^{-1}$, but the MCMC $K$ distributions for the three actual planets do not change beyond their original 1$\sigma$ confidence intervals (Figures \ref{fig:radvel2_1} and \ref{fig:radvel2_2}). The jitter term is similarly unaffected by the extra sinusoidal component. The largest change is with the addition of a 30 day periodic signal with a semi-amplitude of 2.05$^{+1.28}_{-1.29}$ m~s$^{-1}$, where the resulting MCMC distribution for $K_b$ peaks at 5.91$^{+1.39}_{-1.34}$ m~s$^{-1}$, or 7.52$^{+1.8}_{-1.7}$ M$_{\oplus}$, a change of 0.6 \Mearth that is well within the posterior's 68\% credible interval. Adding this fourth 30 day periodic signal decreased the reduced $\chi^2$ value slightly (1.28 to 1.19) and increased the Bayesian Information Criterion (BIC) value slightly (from 225 to 227). In every other case (15d, 16d, 17d, 29d, 31d), both the $\chi^2$ value and BIC value increased with the addition of a fourth signal.

%\newpage

%\newpage 

In both the SYSTEMIC and \textsc{RadVel} analyses, the PFS data are able to place fair constraints on the mass and thus density of planet b. %The average of the SYSTEMIC and \textsc{RadVel} best-fit values for planet b's mass is 7.90 M$_{\oplus}$, while the average of their median values is 7.83 M$_{\oplus}$. Accounting for 1$\sigma$ errors, these averages can span 7.10 to 8.70 M$_{\oplus}$ and 6.30 to 9.37 M$_{\oplus}$. 
These mass ranges overlap with \cite{wolfgang2016}'s predicted mass of 5.1$\pm$1.9 M$_{\oplus}$ (again, representing the astrophysical variation at a planet radius of 1.64 R$_{\oplus}$), but fall on the high side. Our results suggest that planet b $\sim$1.8$\times$ the density of Earth.

\subsection{Gaussian Process Analysis}\label{GPR}
Magnetically active regions (spots, faculae and plages) on a star move into and out of the view of the observer due to stellar rotation. This motion often produces a sinusoidal-like rotational modulation in the observed light curve, and is clearly seen in the {\it K2} light curve of GJ 9827 \citep{niraula2017,rodriguez2017}. This rotational motion of the active regions is also a source of correlated noise in the measured radial velocities \citep{Aigrain2012}. In this section, we investigate, in the framework of Gaussian Process Regression, how the correlated noise due to stellar activity may affect the mass constraints of the planets in GJ 9827.

We followed the procedure described in \cite{dai2017}. In short, we used a Gaussian Process specified by a quasi-periodic kernel to model the correlated stellar noise. The quasi-periodic kernel has the following hyperparameters: the overall covariance amplitude $h$, the relative contribution $\Gamma$ between the squared exponential and periodic part of the kernel (if $\Gamma<<1$ the exponential term dominates the covariance), the correlation timescale $\tau$ of the exponential term, and the period $T$ of the sinusoidal term. As described below, a term for the typical white noise ``stellar jitter" is also included in the covariance matrix along the diagonal, in addition to the RV measurement uncertainties.  Given their common physical origin, we assumed that the rotational modulation seen in the {\it K2} light curve shares the same statistical properties as the correlated stellar noise in the RV dataset. Therefore, we first trained our GP model using the {\it K2} light curve, which is a better sampled and better constrained data set than the PFS RV measurements. We imposed Jeffrey's prior on the various hyperparameters and we sampled the posterior distribution using the affine-invariant MCMC implemented in the code {\tt emcee} \citep{emcee}. The resultant posterior distributions of the hyperparameters ($\tau$ = $6.1^{+4.0}_{-2.3}$ days,  $\Gamma$ = $0.77^{+0.47}_{-0.29}$ and $T = 15.3^{+1.6}_{-1.5}$ days) were then used as priors in the RV analysis.

\begin{figure}[t]
\center
\includegraphics[width=0.8\textwidth]{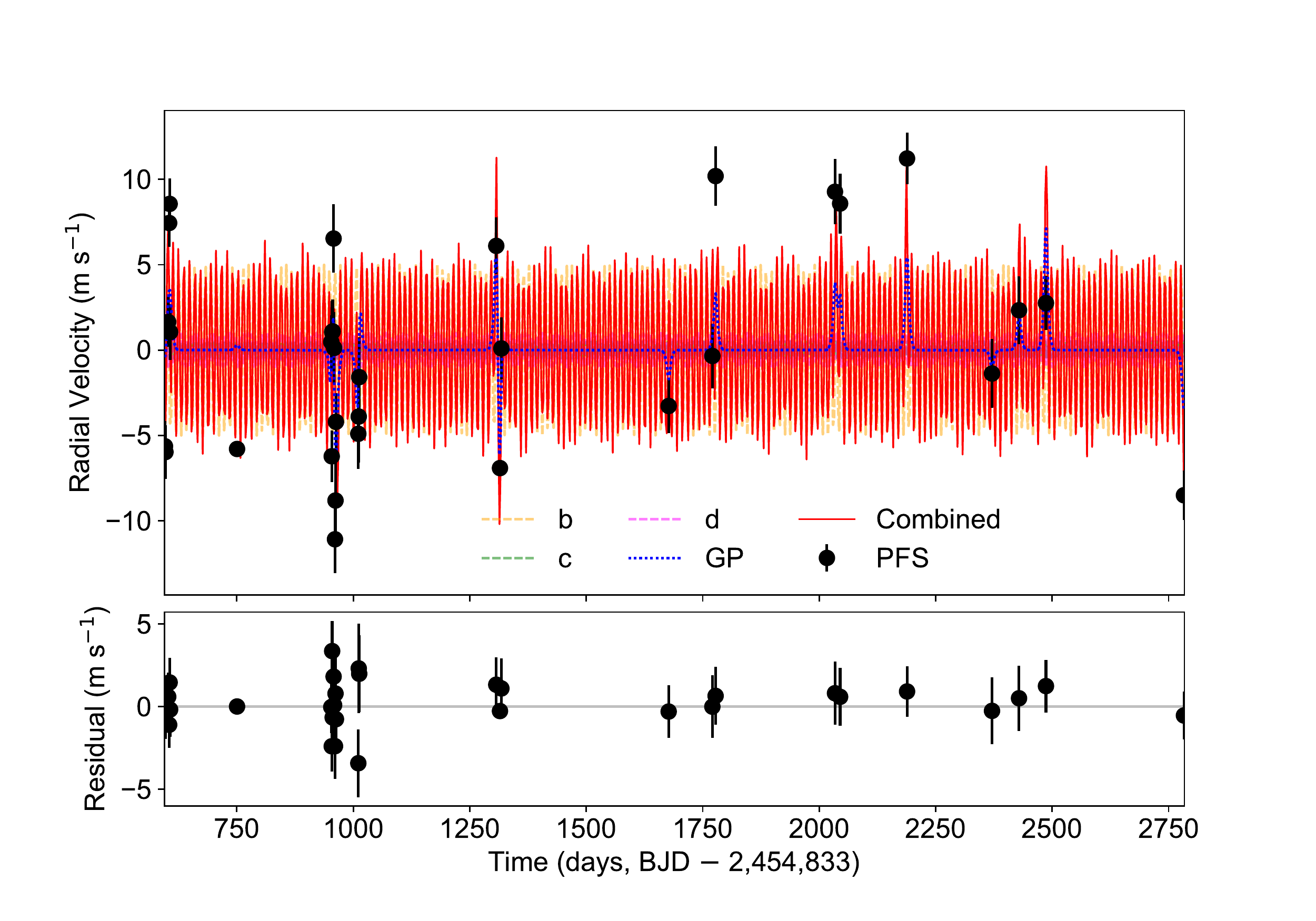} 
\caption{The radial velocity measurements of GJ 9827 from PFS. The colored dashed lines are circular Keplerian models of the three planets. The blue dotted line shows the Gaussian Process model of correlated stellar noise, with priors inferred from the {\it K2} light curve. The red solid line is the combined signal. The Gaussian Process model is poorly constrained (notice the spiky features) when the data have poor temporal sampling. \label{fig:GP_1}}
\end{figure}

\begin{figure}[t]
\center
\includegraphics[width=0.5\textwidth]{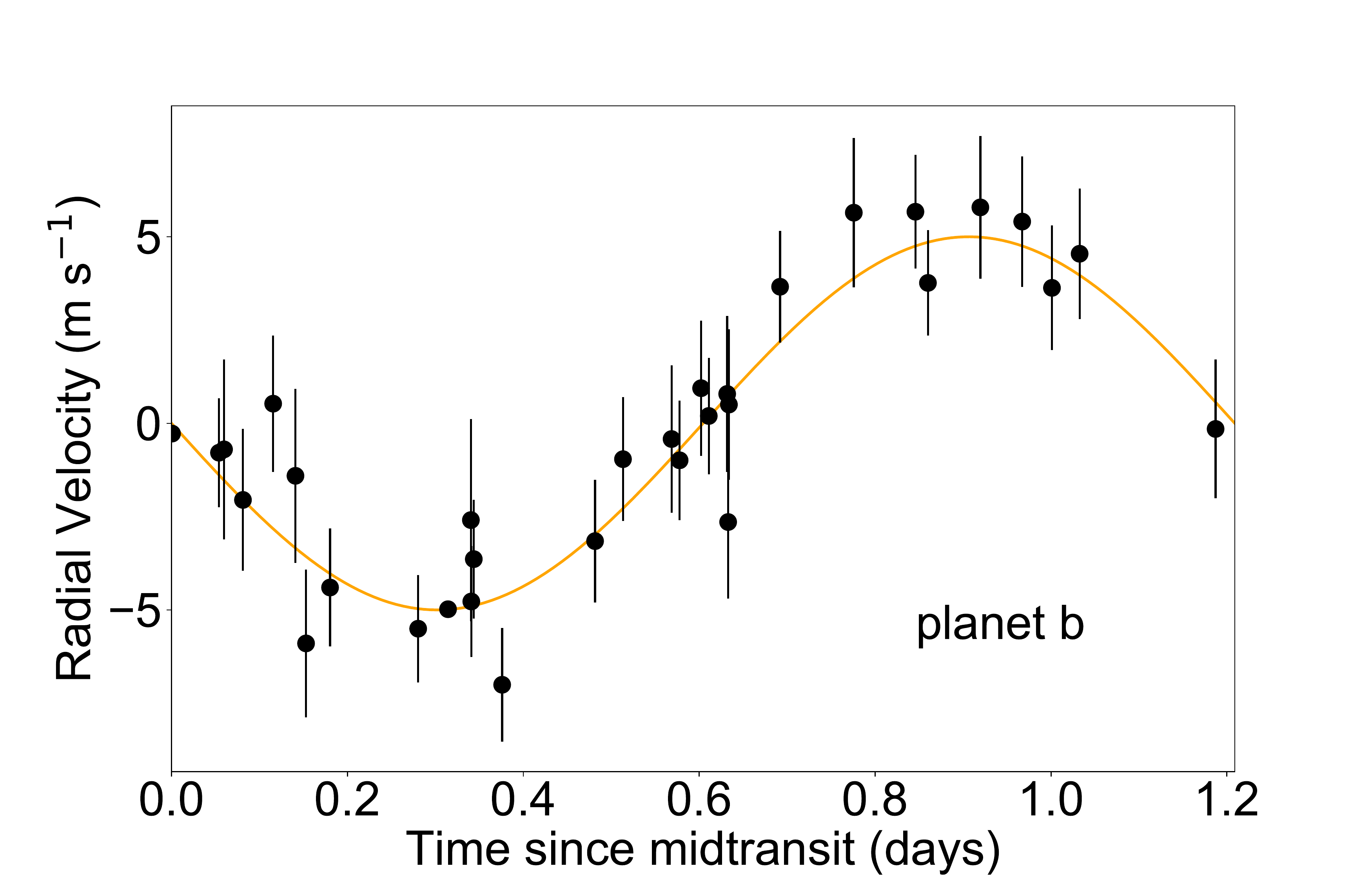} 
\includegraphics[width=0.5\textwidth]{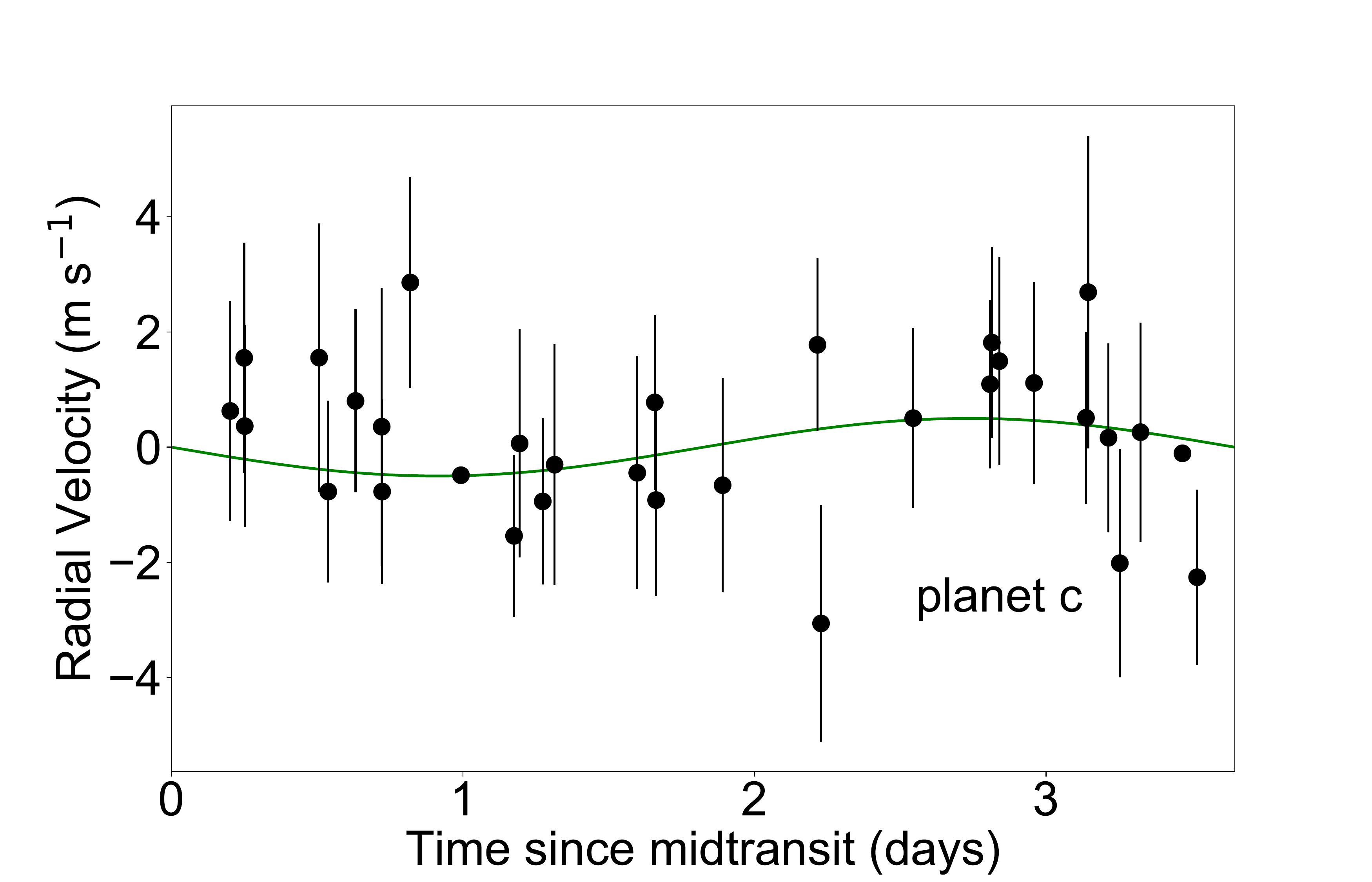} 
\includegraphics[width=0.5\textwidth]{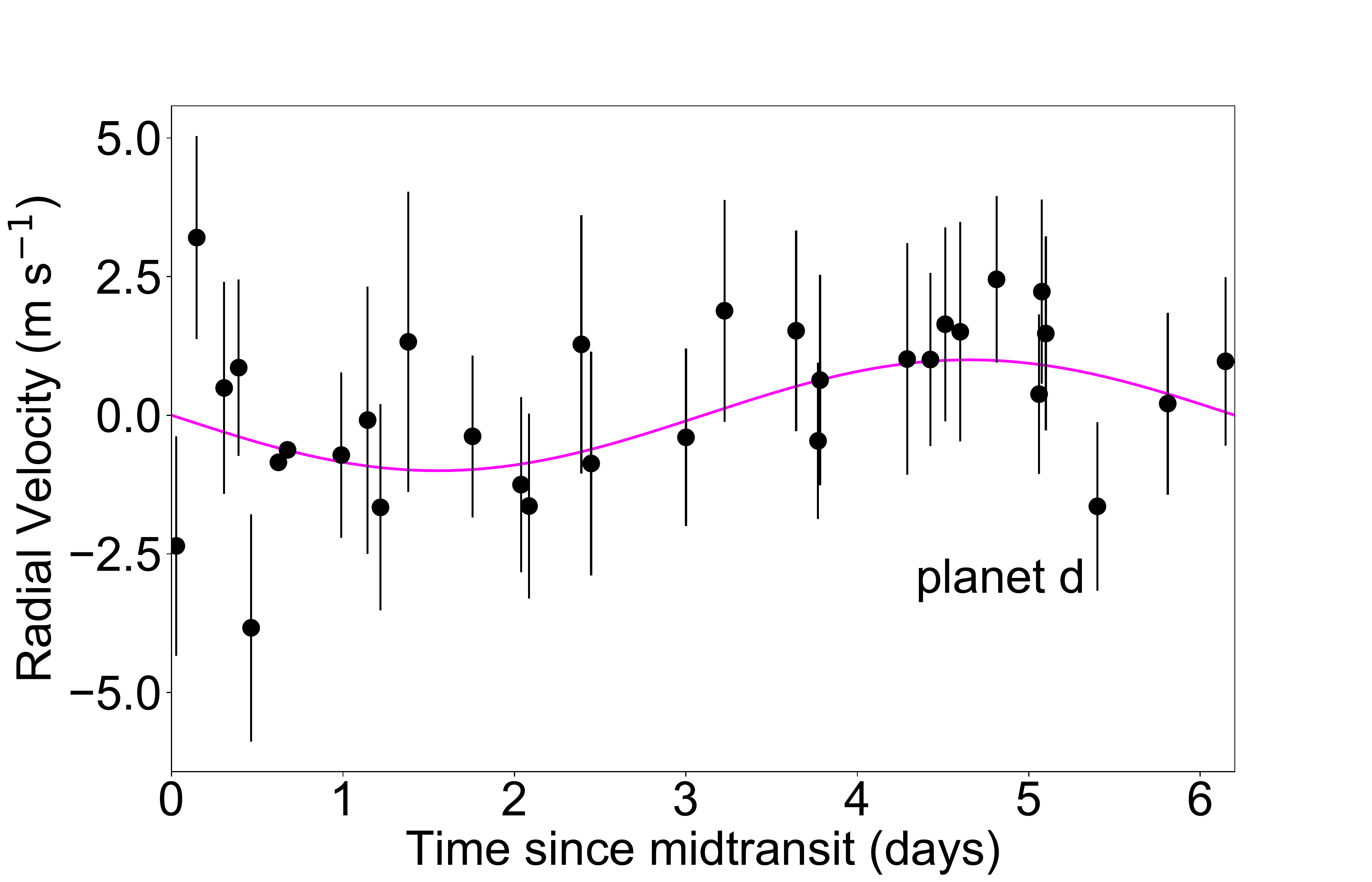} 
\caption{Radial velocity variation phased folded by orbital periods of the planets. The correlated stellar noise was removed by the Gaussian Process analysis. \label{fig:GP_2}}
\end{figure}

In the RV analysis, we included a circular Keplerian signal for each of the three planets. The Keplerian signal was specified by the RV semi-amplitude $K$, the orbital period $P_{\text{orb}}$ and the time of conjunction $t_{\text{c}}$. We included a constant offset $\gamma$ and a jitter parameter $\sigma$ to subsume any instrumental and astrophysical white noise. We imposed Gaussian priors on $P_{\text{orb}}$ and $t_{\text{c}}$ with values derived from {\it K2} transit modeling \citet{rodriguez2017}. We imposed a uniform prior on $\gamma$ and Jeffreys priors on the rest of the parameters. After running a MCMC, we found that following constraints on the RV semi-amplitudes: $K_b = 5.0^{+1.1}_{-1.4}$ m~s$^{-1}$, $K_c < 2.8$ m~s$^{-1}$ and $K_d <3.5 $ m~s$^{-1}$ both at a 95\% confidence level. In \hyperref[fig:GP_1]{Figure 8} we show the PFS-measured RV variation of GJ 9827 (black circles) and the total modeled RV signal (red) separated into the three planet components (orange, green, magenta) and the correlated stellar noise (blue). The best-fit, phase-folded Keplerians for each planet as derived from this analysis, after removal of the correlation stellar noise signal, are shown in \hyperref[fig:GP_2]{Figure 9}. 

The $K_b$ value derived from the GP analysis is lower than that derived in both the SYSTEMIC and \textsc{RadVel} analyses, although overlaps within errors. This seems to suggest that the correlated stellar noise is biasing these programs' inferred RV semi-amplitudes. However, a more careful scrutiny of the PFS RV dataset calls into question the applicability of the GP model. The PFS data were taken prior to the release of {\it K2} observations, so their cadence was not optimized for the orbital periods of the planets nor the stellar rotation period of GJ 9827 revealed by {\it K2} (31$\pm$1 days, although the true rotation period could be half this, $\sim$16 days, \citealt{rodriguez2017}). Many of the PFS observations were separated by weeks or even months, whereas the correlated stellar noise evolves on faster timescale with $\tau$ = $6.1^{+4.0}_{-2.3}$ days and $T = 15.3^{+1.6}_{-1.5}$ days. Without appropriately sampled observations, the GP model for the correlated stellar noise is not well constrained and has the tendency to over-fit the observed RVs. Moreover, given the six year separation between the PFS observations and the {\it K2} observations, the properties of the stellar magnetic active regions might have changed significantly. We therefore adopt the mass constraints from \textsc{RadVel} in the subsequent analysis and discussion. The poor performance of a GP model in this case highlights that such an analysis depends on observation strategy, i.e. high cadence observations with good phase coverage for the planets as well as the variation due to stellar activity.

%%%%SECTION IV: Discussion \& Conclusions %%%%
\section{Discussion \& Conclusions}

\begin{figure*}[t]
\centering
\includegraphics[width=0.68\textwidth]{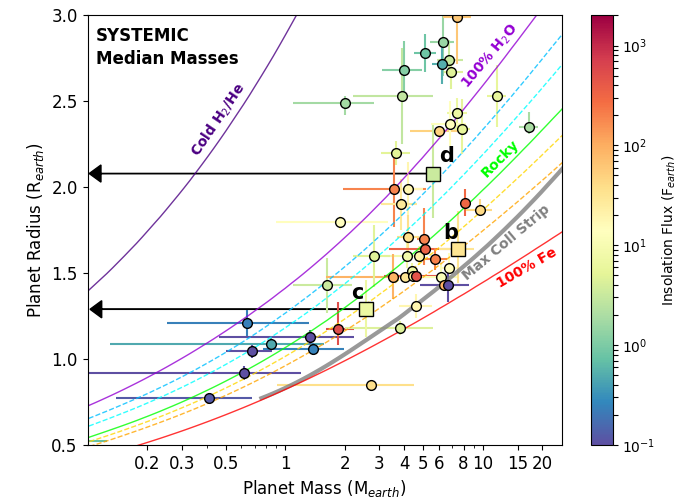}
\includegraphics[width=0.68\textwidth]{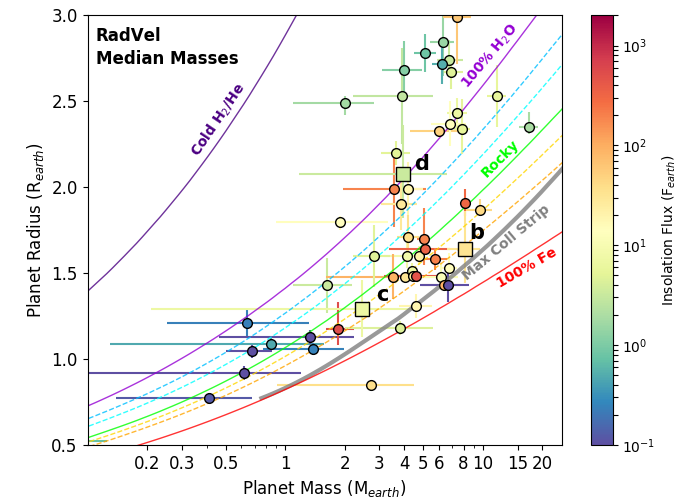}
\caption{A comparison of the median masses and errors for each of the GJ 9827 planets (filled squares), as determined from the MCMC error estimation in SYSTEMIC (top, with upper limits for planet c and d) and in \textsc{RadVel} (bottom), versus other small exoplanets (filled circles) with mass and radius uncertainties $\leq 1.9$ M$_{\oplus}$ and $<1$ R$_{\oplus}$, respectively, from the NASA Exoplanet Archive. The point color corresponds to the planet's insolation flux. Over-plotted in colored solid and dashed lines are planet composition estimates from \cite{zeng2016}. From top to bottom, the compositions are 100\% Fe (solid red), sizes/masses of planets that result from the maximum collisional mantle stripping (solid grey, extrapolated from \cite{marcus2010}), 50\% Fe (dashed orange), 25\% Fe (dashed yellow), MgSiO$_3$ (rocky, solid green), 25\% H$_2$O (dashed cyan), 50\% H$_2$O (dashed blue), 100\% H$_2$O (solid light purple), and cold H$_2$/He (solid dark purple). \label{fig:masscomp}}
\end{figure*}

Based on the planetary mass measurements presented here it appears, unsurprisingly, that planet d likely has a significant volatile envelope, perhaps with a water mass fraction between 50-100\%. Both the SYSTEMIC and \textsc{RadVel} masses also allow for lighter, perhaps H$_2$/He-dominated envelope. Being the farthest away, planet d also has the lowest insolation flux, $\sim30\times$ that of the Earth, and is thus the least susceptible to photoevaporation. It falls towards the lower half of the second peak, centered around 2.4 R$_{\oplus}$ with a spread from $\sim 1.8 - 3.5$ R$_{\oplus}$, in the exoplanet radius distribution found by \citet{fulton2017}.

The mass of planet c is the least constrained due to its small size and resulting small RV signal ($\sim$1-1.5 m~s$^{-1}$). The composition of planet c as inferred from  \hyperref[fig:masscomp]{Figure 10} ranges from $\gtrsim$50\% Fe by mass to $\gtrsim$50\% water by mass. If planet c does have a significant volatile envelope, it would be unusual in the picture of super-Earths described in \S1, since its radius is below the predicted 1.6 R$_{\oplus}$ cut-off between rocky and gas-enveloped planets.

%\newpage
\subsection{High Density of GJ 9827b}

The most significant mass constraints occur for planet b; they also happen to be the most astrophysically interesting.
At $1.64^{+0.22}_{-0.20}$ \Rearth, GJ 9827 b straddles the rocky transition radius found by \citet{rogers2015}: above this transition most planets are not rocky, but below planets could be either rocky or gaseous.  Planet b also falls at the lower edge of the gap in the \emph{Kepler} radius distribution found by \citet{fulton2017}.  If this gap is caused by photoevaporation \citep{lopez2016,owen&wu2017} as is suggested by the period dependence of the gap \citep{vaneylen2017}, one would expect planet b to be rocky.

This expectation is borne out by the radial velocity observations presented here: at $\sim 8 \pm 2$ \Mearth, GJ 9827 b is squarely in the rocky regime, with the best fit indicating an iron mass fraction of $\gtrsim$50\%. In fact, this planet is one of the densest planets at its respective mass and radius, and is one of the most massive terrestrial planets found yet.  With the observed super-Earth population as context for our surprising mass measurement, we therefore ask: just how anomalous is GJ 9827 b?

From an empirical perspective, GJ 9827 b is much denser than the average 1.6 \Rearth planet, but not unexpectedly so.  Comparing the measured mass to probabilistic mass-radius (M-R) relations, which characterize this diversity, affords us a quantitative answer to this question.  \citet{wolfgang2016} predicts that planet b would have a mass of 5.1$\pm$1.9 \Mearth, where the ``error" actually represents the astrophysical standard deviation in the masses of planets with radii $< 4$ \Rearth.  Therefore, planet b's mass is $\sim 1.5\sigma$ above the average mass for planets with 1.6 \Rearth ($\sim 90\%$ of all planets at that size would be less massive than GJ 9827 b).  On the other hand, the mass prediction reported in \citet{rodriguez2017} based on the M-R relation of \citet{chen2017}, predicts $3.5^{+1.4}_{-0.9}$ \Mearth, which would place planet b's measured mass in the 99.8th percentile ($> 3\sigma$) and would suggest the planet is much more unusual.  

Before declaring GJ 9827 b as the next extreme super-Earth (significantly more massive than other planets of similar size), however, it is worth noting three things.  First, the super-Earth transition region in the \citet{chen2017} M-R relation was constrained on the low-mass end with Solar System terrestrial bodies, and so could be underestimating the astrophysical scatter in exoplanet super-Earths.  Second, it is much more difficult to obtain statistically significant constraints on the mass of a 2 \Mearth, 1.6 \Rearth planet than it is for a 8 \Mearth, 1.6 \Rearth planet, and so there are fewer reported low-mass planets than high-mass planets at any given super-Earth radius.  This bias is an unavoidable aspect of the datasets which are used to construct these probabilistic M-R relations, and thus has likely caused both \citet{wolfgang2016} and \citet{chen2017} to underestimate the actual astrophysical scatter in the population.  

Third, there is still a question of how much individual mass measurements depend on the details of the observations and on the radial velocity model used to analyze them. The statistically significant disagreement in mass for Kepler-10c between HARPS (\citealt{dumusque2014}, plotted in \hyperref[fig:masscomp]{Figure 10} as the green point all the way on the right) and HIRES \citep{weiss2016} serves as a warning for planets whose masses are unusually high. As more data inevitably arrives to constrain the GJ 9827 system, we will need to ensure that suboptimal sampling and incomplete model choice have not colluded to produce an anomalous mass measurement \citep{rajpaul2017}. As these new data are presented, we will also need to assess what kinds of stellar activity models are appropriate.  As is discussed in \S \ref{GPR}, the answer to this question will likely depend on both the dataset and the details of the RV model.  Moving forward, it is becoming increasingly prudent to test one's analysis model with simulated data, which can determine the degree to which one's stellar activity model over-fits the planetary signal (i.e. \citealt{DEJones2017}).

%The radius of planet b is on the predicted boundary of rocky versus enveloped by volatiles, but its mass places it squarely in the rocky regime, with the best fit indicating an iron mass fraction of $\gtrsim$50\%. 
From a theoretical perspective, planet b is certainly intriguing. Given its possibly high iron fraction, it is among the handful of planets that may have undergone mantle stripping as a result of giant impacts. The period and density of planet b place it outside the range of tidal disruption per the relations of \citet{raymond2013}, but it may still have had a violent past that influenced its present radius. Significant brightening and variability observed in emission from debris disks have been suggested as evidence of collisions between large bodies in the era of small planet formation \citep{meng2014}. Simulations also indicate that giant impacts occur in the later stages of planet formation \citep[see][and references therein]{wyatt2016, Quintana2016}, although high impact velocities and impact angles between 0-30$^{\circ}$ are required to produce the maximally effective collisions for stripping mantles \citep{agnorasphaug2004,asphaug2006,asphaug2009,marcus2009}. Specifically, to reproduce the current mass and high iron mass fraction of planet b, \citet{marcus2010} predict that a collision with an object of similar mass at $\sim$60-80 km~s$^{-1}$ is required. However, the metallicity of GJ 9827 (here derived to be [Fe/H]$\sim$-0.3) suggests that its natal disk would have a lower solid surface density, so might be less likely to produce a large number of super-Earth-sized bodies \citep[e.g.,][]{dawson2015}. 
%Can we say anything about likelihood of impacts due to the resonance of planet b with planet c but not with planet d? 

Whatever planet b's history, it appears to be one of the most massive super-Earths detected to date, similar in mass to 55 Cnc e and K-20b but slightly smaller in radius. \citet{demory2016} used \textit{Spitzer}/IRAC phase curve observations of 55 Cnc e to measure the longitudinal thermal brightness variations, and found a maximum hemisphere-averaged temperature of 2700$\pm$270 K, which is slightly higher than the highest possible equilibrium temperature. The authors also found a large day-to-night temperature contrast, indicating little to no atmospheric circulation on 55 Cnc e, but did observe a hot spot offset from the sub-stellar point. They  suggest one explanation of the data is that 55 Cnc e has no atmosphere and a low albedo, and that the hot spot offset is driven by an eastward flow of molten lava ($\sim$2700 K is hot enough for silicate rocks to be molten) on the planet's day side surface. Similar phase curve variations of GJ 9827b would provide an illuminating comparison for an even denser, shorter period planet.

\begin{figure*}[t]
\centering
\includegraphics[width=1\textwidth]{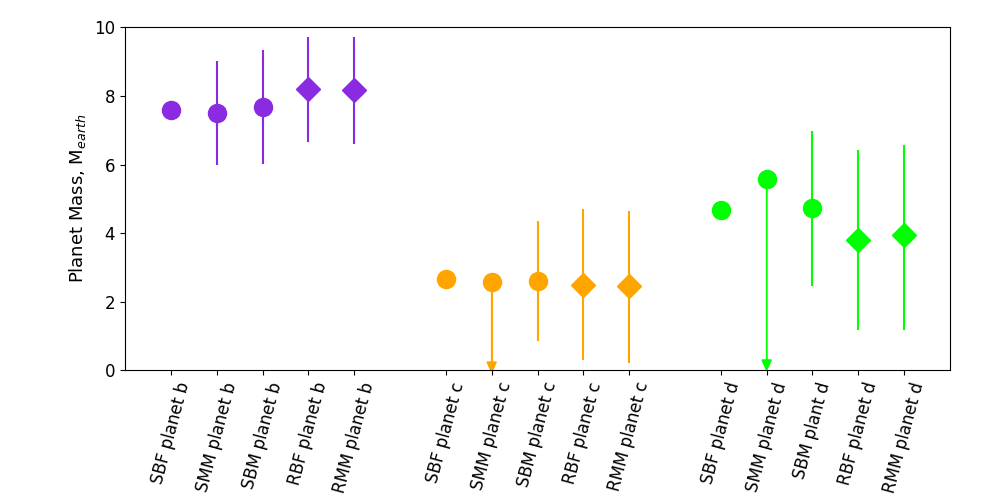}
\caption{A comparison of the masses for each of the GJ 9827 planets, as determined from SYSTEMIC (S, circles) and \textsc{RadVel} (R, diamonds) best-fits (BF), MCMC medians (MM), and (for SYSTEMIC) bootstrap medians (BM). The results of planet b are plotted in purple, planet c in orange, and planet d in green. %The dashed lines across each set of four points indicate the (unweighted) averages of the best-fit values, and the dashed-dotted lines indicate the (unweighted) averages of the median values (both bootstrap and MCMC). 
The error bars on each point correspond to the 1$\sigma$ errors; SYSTEMIC does not report errors on the best-fit parameters. The MCMC results from SYSTEMIC are plotted with arrows to represent upper limits.\label{fig:masscomp2}}
\end{figure*}

\subsection{SYSTEMIC vs. RadVel Results}
As far as we know, this is the first publication to compare the RV fitting results of SYSTEMIC and \textsc{RadVel}. In this case the only fitted parameter is the RV semi-amplitude, significantly reducing the complexity of the analysis, but the comparison of results still has implications for determining the masses of planets detected via the \textit{Kepler}, \textit{K2}, and soon \textit{TESS} missions. That is, using the same data and the same assumptions, do different (relatively) automated analysis programs give significantly different results? Looking at \hyperref[fig:masscomp2]{Figure 11}, the answer in this case appears to be no -- while the full ranges of derived planet masses from SYSTEMIC and \textsc{RadVel} indicate that the mass of planet b could be between $\sim$6 and 9.5 M$_{\oplus}$, the best-fit and MCMC median values differ by 0.60 and 0.66 M$_{\oplus}$, respectively. %; the bootstrap median SYSTEMIC value and the \textsc{RadVel} MCMC median values differ by 0.49 M$_{\oplus}$. 
These differences are less than the respective errors from each program ($\sim$1.5 M$_{\oplus}$). For the other two planets, the differences between SYSTEMIC and \textsc{RadVel} derived masses are, in M$_{\oplus}$: -0.15 (planet c, best fit), -0.11 (planet c, SYSTEMIC MCMC upper limit vs. \textsc{RadVel} MCMC median), %-0.15 (planet c, SYSTEMIC bootstrap vs. \textsc{RadVel} MCMC), 
0.87 (planet d, best fit), and -1.65 (planet d, SYSTEMIC MCMC upper limit vs. \textsc{RadVel} MCMC median). %, and 0.80 (planet d, SYSTEMIC bootstrap vs. \textsc{RadVel} MCMC). 
However, as noted above the mass constraints on planet c and d's masses are not as good, and the SYSTEMIC values are really upper limits. The assumed prior distribution has a strong effect on the shape of the posterior for these two planets: SYSTEMIC's log-uniform prior on mass produces a notably different posterior shape than \textsc{RadVel}'s uniform prior on $K$. Therefore, while we are reporting the median and central 68\% credible interval on the \textsc{Radvel} planet masses, the user should consider which prior distribution better reflects their expectations about planet mass pre-measurement and be aware that this data provides only weak constraints on the masses of planets c and d.
An interesting next step would be to test the differences between SYTEMIC and \textsc{RadVel} on systems with a larger/fewer number of observations, data from different telescopes, and/or stronger planetary RV signals.

\subsection{Conclusions}

\citet{niraula2017} and \citet{rodriguez2017} recently reported the \textit{K2} detection of three short period super-Earth exoplanets around the nearby K dwarf GJ 9827. Here we present radial velocity observations of the star from the Planet Finder Spectrograph on Magellan II, spanning from 2010 to 2016. We analyze these observations using two separate RV packages, SYSTEMIC and \textsc{RadVel}, to derive mass constraints on the planets. We also tried analyzing the observations using Gaussian Process regression with a quasi-periodic kernel. However, the RV observations in this case are sparsely sampled and thus the GP model is not well constrained. Furthermore, the hyperparameter priors are derived from the {\it K2} light curve observed in 2017, whereas the RV observations span 2010-2016 and could be affected by significantly different magnetically active regions on the surface of GJ 9827.

Based on the SYSTEMIC and \textsc{RadVel} analyses, we find planet b is $\sim$8 M$_{\oplus}$, planet c is $\lesssim$2.5 M$_{\oplus}$ although with uncertainties between $\sim$70\% and $>100$\%, and planet d is $\lesssim$6 M$_{\oplus}$, also with large uncertainties. These mass constraints suggest that planet b is $\gtrsim$50\% iron by mass and planet d has a significant volatile envelope, but do not place strong constraints on the composition of planet c. Further RV observations of the system may help pin down the masses and thus compositions of the planets in this system, which are of special interest because their radii span the predicted and recently detected radius gap in $P<100$ small planets detected by \textit{Kepler}.

% -
\acknowledgments
J.K.T. thanks Thomas Connor for his help in making python plots with error bars \textit{and} a color bar. A.W. acknowledges support by the National Science Foundation under Award No. 1501440. We thank the referee for their very thoughtful and helpful comments that provided in-depth knowledge and improved the manuscript.  

This research has made use of the NASA Exoplanet Archive, which is operated by the California Institute of Technology, under contract with the National Aeronautics and Space Administration under the Exoplanet Exploration Program. This research has also made use of the Exoplanet Follow-up Observation Program website, which is operated by the California Institute of Technology, under contract with the National Aeronautics and Space Administration under the Exoplanet Exploration Program. 

%\facility{facility ID}
\facilities{Magellan:Clay (Planet Finder Spectrograph)} 
\software{Systemic (\url{https://github.com/stefano-meschiari/Systemic2}), Astropy (\url{http://dx.doi.org/10.1051/0004-6361/201322068}), Matplotlib (\url{http://dx.doi.org/10.1109/MCSE.2007.55}), RadVel (\url{https://github.com/California-Planet-Search/radvel/tree/v1.0.3})}

\bibliographystyle{yahapj}
\bibliography{references}

%\appendix
%\section{appendix section}

\end{document}